\documentclass[aps,prb,superscriptaddress,preprint]{revtex4-1}
\usepackage[utf8]{inputenc}

\usepackage{times} \usepackage{amsmath} \usepackage{amsfonts}
\usepackage{amssymb} \usepackage{epsfig} \usepackage{siunitx}
\usepackage{float}
\usepackage{ifthen}
\usepackage{xspace}
\usepackage{relsize}
\usepackage{algorithm}
\usepackage{algorithmic}

\newcommand{\md}{\mathrm{d}} \newcommand{\me}{\mathrm{e}}
\newcommand{\mi}{\mathrm{i}}

\usepackage{todonotes}

\begin{document}

\title{Quantum tomography of electrical currents}

\author{R.  Bisognin$^{1 \dag}$, A.  Marguerite$^{1 \dag}$, B.  Roussel$^{2,5}$,
M.  Kumar$^{1}$, C.  Cabart$^{2}$, C.  Chapdelaine$^{4}$, A.
Mohammad-Djafari$^{4}$, J.-M.  Berroir$^{1 }$, E.  Bocquillon$^{1 }$, B.
Pla\c{c}ais$^{1 }$, A.  Cavanna$^{3}$, U.  Gennser$^{3 }$, Y.  Jin$^{3 }$, P.
Degiovanni$^{2}$, and G.  F\`{e}ve$^{1 \ast}$ \\
\normalsize{$^{1}$ Laboratoire
de Physique de l\textquoteright Ecole normale sup\'erieure, ENS, Universit\'e
PSL, CNRS, Sorbonne Universit\'e, Universit\'e Paris-Diderot, Sorbonne Paris
Cit\'e, Paris, France}\\
\normalsize{$^{2}$ Univ Lyon, Ens de Lyon, Universit\'e
Claude Bernard Lyon 1, CNRS,}\\
\normalsize{Laboratoire de Physique, F-69342
Lyon, France.}\\
\normalsize{$^{3}$ Centre de Nanosciences et de
Nanotechnologies (C2N), CNRS, Univ.  Paris-Sud, Universit\'{e} Paris-Saclay, 91120 Palaiseau, France.}\\
\normalsize{$^{4}$
Laboratoire des signaux et syst\`{e}mes, CNRS, Centrale-Sup\'{e}lec-
Universit\'{e} Paris-Saclay.}\\
\normalsize{$^{5}$
European Space Agency - Advanced Concepts Team, ESTEC, Keplerlaan 1, 2201 AZ Noordwijk, The Netherlands.}\\
\normalsize{$^\ast$ To whom correspondence
should be addressed; E-mail: gwendal.feve@ens.fr.}\\
\normalsize{$^\dag$ These
authors contributed equally.}\\ }

\begin{abstract}
	In quantum nanoelectronics, time-dependent electrical currents
are built from few elementary excitations emitted with well-defined
wavefunctions.  However, despite the realization of sources generating quantized numbers of excitations, and despite
the development of the theoretical framework of time-dependent quantum
electronics, extracting electron and hole wavefunctions from electrical
currents has so far remained out of reach, both at the theoretical and experimental
levels. In this work, we demonstrate a quantum tomography protocol which extracts the
generated electron and hole wavefunctions and their emission probabilities from
any electrical current. It combines two-particle interferometry with signal
processing. Using our technique, we extract the wavefunctions
generated by trains of Lorentzian pulses carrying one or two electrons.
By demonstrating the synthesis and complete characterization of
electronic wavefunctions in conductors, this work offers perspectives for
quantum information processing with electrical currents and for investigating
basic quantum physics in many-body systems.
\end{abstract}

\maketitle

In the field of quantum technologies, controlling elementary excitations such as
single photons \cite{singlephoton}, single atoms \cite{atombyatom} or single
ions \cite{singleion} is a resource for encoding quantum
information as well as a way to develop our understanding of basic quantum
physics in complex many-body problems.  In quantum electronics, the availability of on-demand single electron
sources\cite{Ondemand,Leicht2011,Dubois2013,Fletcher2013} offers the possibility
to generate time-dependent electrical currents carrying a controlled number of
electron and hole excitations of a degenerate electronic fluid.  At low
temperatures, phase coherence is preserved, such that describing these
excitations in terms of well-defined wavefunctions is meaningful.  Additionally,
by implementing electron sources in ballistic low-dimensional conductors
\cite{Roussely2018,Ondemand,Fletcher2013}, the elementary excitations can be
guided along one-dimensional channels and used as flying qubits
\cite{Bertoni2000,Yamamoto2012} carrying information encoded in their quantum
state.  However, despite the development of a rich experimental toolbox to
generate and propagate electronic states in a controlled way, very few tools are
currently available to characterize these states.  In particular, measuring
electron or hole wavefunctions embedded within a quantum electrical current has,
so far, been out of reach.

\begin{figure}[h!]
	\includegraphics[width=\linewidth]{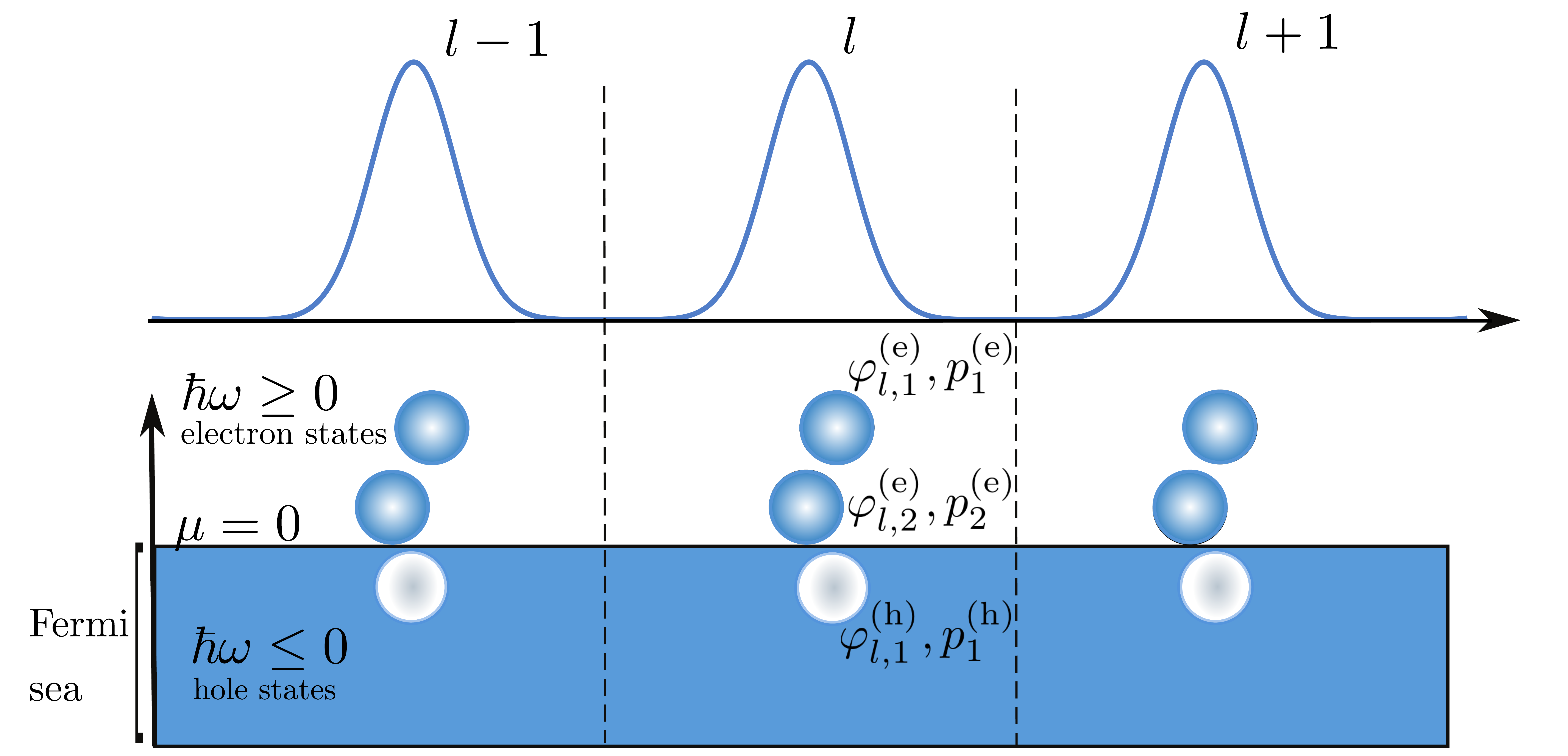}
	\caption{
Sketch of the elementary excitations generated by a periodic current $I(t)$.
Electron and hole states are generated at each period $l$ above and below the
Fermi sea.  The time translated wavefunctions
$\varphi^{(\alpha)}_{l,i}(t)=\varphi^{(\alpha)}_i(t-lT)$, for $\alpha=\text{e}$
(electron) or $\text{h}$ (hole) and $1 \leq i \leq N_{\alpha}$ are emitted at each
period with probability $p^{(\alpha)}_{i}$.  }
\label{fig1}
\end{figure}

This absence of a universal tomography protocol in the fermionic case may seem
peculiar, considering that such protocols are now commonly implemented to
reconstruct the state of bosonic fields \cite{Smithey:1993-1,Lvovsky:2009-1}.
However, there are important differences between bosonic and fermionic fields.
Firstly, bosonic tomography protocols involve the use of a classical field
\cite{Lvovsky:2009-1} which has no counterpart for fermions.  Secondly, the
vacuum of a fermionic system being a Fermi sea, the electron and hole excitations
are thus defined by the addition and removal of a particle.  Quantum state
reconstruction in the fermionic case can be illustrated by the sketch of
Fig.~\ref{fig1}.  In a one-dimensional conductor, a $T$-periodic source generates
a time-dependent current consisting of periodic pulses labeled by the
index $l\in\mathbb{Z}$.
To define unambiguously the electron and hole excitations, we take the conductor
at chemical potential $\mu=0$ and temperature $T_{\text{el}}=$\SI{0}{\kelvin} as
a reference.  The electronic excitations correspond to the filling of the states
above the Fermi sea (energy $\hbar \omega \geq 0$) and the hole excitations to
the emptying of the states below the Fermi sea ($\hbar \omega \leq 0$). We
introduce in Fig.~\ref{fig1} the emitted time-translated electron $(\text{e})$ and hole
$(\text{h})$ wavefunctions $\varphi^{(\alpha)}_{l,i}(t)=\varphi^{(\alpha)}_i(t-lT)$, and
their emission probabilities $p^{(\alpha)}_{i}$ where $\alpha = \text{e}$ or $\text{h}$ labels the
electron or hole states and $i$ runs from $1$ to $N_{\alpha}$, the total number
of electron ($N_\text{e}$) and hole ($N_\text{h}$) wavefunctions emitted per period. These
emitted electron and hole wavefunctions form a set of mutually orthogonal
states: $\langle \varphi^{(\alpha')}_{l',i'}|\varphi^{(\alpha)}_{l,i}\rangle
=\delta_{i,i'}\,\delta_{\alpha,\alpha'}\,\delta_{l,l'}$.

By combining two-particle interferometry\cite{Marguerite2017} with signal
processing \cite{These:Roussel}, we demonstrate here a quantum current analyzer
which extracts the emitted wavefunctions $\varphi^{(\alpha)}_{l,i}(t)$ and their
emission probabilities $p^{(\alpha)}_{i}$ from any periodic electrical current.
For benchmarking, we first apply our analyzer on sinusoidal currents to validate
our extraction method in the general case, when several excitations are emitted
with non-unit probability.  Sinusoidal drives are well suited to test the
robustness of our procedure by comparing our results with parameter free
theoretical predictions.  We then apply our technique to trains of Lorentzian
pulses carrying an integer charge $q=-\text{e}$ and $q=-2\text{e}$ and extract their full content in
terms of single electron wavefunctions. At zero temperature, Lorentzian pulses of
integer charge $-\text{e} N_\text{e}$ are predicted to generate an integer number $N_\text{e}$ of excitations exclusively above the Fermi
sea\cite{Levitov96,Keeling06,Dubois2013,Gabelli2013}.
By extracting all the emitted wavefunctions, we observe that
thermal effects lead to the generation of a statistical mixture between the expected zero temperatures wavefunctions and additional undesired states. From the measurement of the emission probability of each generated wavefunction,
we provide a quantitative analysis of the purity of the generated electronic states.

By identifying specific single electron and hole wavefunctions and determining
their emission probabilities for various types of time dependent currents, our
work opens the way to a precise and systematic characterization of quantum
information carried by electrical currents.

\section*{Results}

{\bf Electronic coherence and Wigner distribution.}
The main difficulty behind the extraction of the electron and hole wavefunctions from an electrical current
lies in the explicit connection between the wavefunctions and a measurable physical quantity. So far, most of
the characterizations of the excitations generated by electronic sources have been limited to the measurements of the average electrical current $I(t)$
\cite{Ondemand,Kataoka2016} and electronic distribution function $f(\omega)$
\cite{Altimiras2009,Fletcher2013}.  They provide information on the time and
energy distributions but cannot access the phase of electronic wavefunctions
which requires the use of interferometry techniques.

In analogy with optics, all interference effects are encoded in the
first-order electronic coherence $\mathcal{G}^{(\text{e})}_{\rho,x}(t,t')$\cite{Bocquillon2014, Haack2013} defined as the
time correlations of the fermion field $\hat{\Psi}(x,t)$ which annihilates an electron at position
$x$ and time $t$ of the one-dimensional conductor:
$\mathcal{G}^{(\text{e})}_{\rho,x}(t,t')=\langle \hat{\Psi}^{\dag}(x,t')
\hat{\Psi}(x,t)\rangle_\rho$. To simplify the notations in the rest of the paper,
we suppress the superscript $(\text{e})$ and the dependence on the position $x$ and on the many-body density operator
$\rho$ in the expression of the electronic coherence which is written as $\mathcal{G}(t,t')$.
More generally, $\mathcal{G}(t,t')$ contains all the information on the
single particle properties of the many-body electronic state.  Electronic
coherence being a priori a complex function, it is more convenient to use the
electronic Wigner distribution \cite{DegioWigner2013,Kashcheyevs2017}
$W(t,\omega)$ obtained from $\mathcal{G}(t+\tau/2,t-\tau/2)$
by Fourier transform
along the time difference $\tau$.  $W(t,\omega)$ is a real function of
marginal distributions $I(t)$ and $f(\omega)$ obtained by respectively integrating
$W(t,\omega)$ over energy $\omega$ and time $t$, thereby demonstrating that they only
provide partial information.

Subtracting the reference contribution characterized by the zero-temperature Fermi
distribution $\Theta(-\omega)$ (where $\Theta$ is the Heaviside
function) defines
$\Delta_0 W(t,\omega)=W(t,\omega)-\Theta(-\omega)$ (or equivalently
its Fourier transform $\Delta_0 \mathcal{G}(t,t')$). $\Delta_0 W(t,\omega)$ and
$\Delta_0 \mathcal{G}(t,t')$ are the key quantities
that we explicitly connect to the wavefunctions $\varphi^{(\alpha)}_{l,i}$ and emission
probabilities $p^{(\alpha)}_{i}$.  This connection is trivial in the pure state single-body
case, that is when a single excitation (either electron or hole) of wavefunction
$\varphi$ is emitted with unit probability. In this simple limit,
$\Delta_0 W(t,\omega) = W_{\varphi}(t,\omega)$, where
$W_{\varphi}(t,\omega)$ is the Wigner representation \cite{Wigner} of the
wavefunction $\varphi$:
\begin{equation}
	W_{\varphi}(t,\omega)=\int \md\tau\,
\varphi\left(t+\frac{\tau}{2}\right) \varphi^{*}\left(t-\frac{\tau}{2}\right)
	\me^{\mi \omega \tau}.
\label{Wigphi}
\end{equation}
In a recent experiment \cite{Jullien2014}, Jullien
\textit{et al.} performed the first reconstruction of $W(t,\omega)$ in the
case of a periodic train of single-electron Lorentzian pulses.  Assuming that
the single-body limit was valid, they extracted the electronic wavefunction
$\varphi$ using $\Delta_0 W(t,\omega) = W_{\varphi}(t,\omega)$.  However,
the single-body limit can never be completely achieved due to the presence of
thermal excitations, to the periodic emission from the source or to
deformations of the current pulse associated with imperfections of the
voltage drive or due to more fundamental effects such as the Coulomb interaction.
Additionally, for multi-electron states, such as Lorentzian pulses carrying an
integer number of excitations
$N_\text{e}>1$\cite{Keeling06,Grenier2013,Ferraro2018}, going beyond the single-body
limit to extract the electronic wavefunctions is absolutely required.

In the more complex case where one has to consider several wavefunctions
generated with arbitrary probabilities, only specific sets of drives
\cite{Vanevic2007,Vanevic2016,Yin2019} have been theoretically investigated.  Furthermore,
the connection between experimentally accessible quantities and the emitted
wavefunctions was missing. Following the work of Ref.
\onlinecite{These:Roussel}, we explicitly connect the emitted electron $\varphi^{(\text{e})}_{l,i}$ and
hole $\varphi^{(\text{h})}_{l,i}$ wavefunctions to the electronic coherence $\Delta_0 \mathcal{G}$ by diagonalizing $\Delta_0 \mathcal{G}$ in the subspace of electron and hole states (see Methods).
As a result of the diagonalization procedure, $\Delta_0 \mathcal{G}$ can be decomposed in the basis of
electron and hole states $\varphi^{(\alpha)}_{l,i}$ by introducing the matrix elements $g^{(\alpha \beta)}_{ij}(l) =  \langle \varphi^{(\alpha)}_{l,i}|\Delta_0 \mathcal{G} |\varphi^{(\beta)}_{0,j}\rangle = \int \md t\, \md t' \varphi^{(\alpha)}_{l,i}(t)^* \Delta_0
	\mathcal{G}(t,t') \varphi^{(\beta)}_{0,j}(t')\,$:

\begin{widetext}
\begin{subequations} \label{eq:final-decomposition}
	\begin{align}
	\Delta_0\mathcal{G}(t,t')
&= \sum_{i =1}^{N_\text{e}} \sum_{(l,l')\in\mathbb{Z}^2} g^{(\text{ee})}_i(l-l')\,
\varphi^{(\text{e})}_{l,i}(t)\varphi^{(\text{e})}_{l',i}(t')^* -\sum_{i =1}^{N_\text{h} }\sum_{(l,l'
\in\mathbb{Z}^2)}
g^{(\text{hh})}_i(l-l')\,\varphi^{(\text{h})}_{l,i}(t)\varphi^{(\text{h})}_{l',i}(t')^* \nonumber \\
	&+ \sum_{i,j}\sum_{(l,l' \in\mathbb{Z}^2)}\; \left(
g^{(\text{eh})}_{ij}(l-l')\,\varphi^{(\text{e})}_{l,i}(t)\varphi^{(\text{h})}_{l',j}(t')^*
		 +g^{(he)}_{ji}(l-l')\,\varphi^{(\text{h})}_{l,j}(t)\varphi^{(\text{e})}_{l',i}(t')^*\right) \tag{\ref{eq:final-decomposition}}
\end{align}
\end{subequations}
\end{widetext}

As the electron and hole wavefunctions $\varphi^{(\alpha)}_{i}$ are
extracted from the diagonalization of $\Delta_0\mathcal{G}$, it naturally implies
that there are no quantum coherences in Eq.~\eqref{eq:final-decomposition} between states  $\varphi^{(\alpha)}_{l,i}$
and $\varphi^{(\alpha)}_{l',i'}$ whenever $i\neq i'$: $g^{(\alpha \alpha )}_{i\neq i'}=0$.

Each term of Eq.~\eqref{eq:final-decomposition} can be separately interpreted.
The first (second) term represents the contribution of electron (hole)
wavepackets to the first-order coherence.  For $l=l'$, the real numbers
$0\leq g^{(\text{ee})}_i(0)\leq 1$ and $0\leq g^{(\text{hh})}_i(0)\leq 1$ represent
the probability for emitting the electron ($\varphi^{(\text{e})}_i$) and hole
($\varphi^{(\text{h})}_i$) wavefunctions. Following the notation introduced at the
beginning of the paper, we thus have
$p^{(\alpha)}_i=g^{(\alpha \alpha)}_i(0)$. Compared to the simple picture sketched
in Fig.~\ref{fig1}, the $T$-periodicity of
the source requires to consider also the complex numbers $g^{(\text{ee})}_i(l-l')$
(resp.  $g^{(\text{hh})}_i(l-l')$) for $l\neq l'$ representing coherences between
electronic (resp. hole) wavepackets emitted at different periods.  The
last two
terms of Eq.~\eqref{eq:final-decomposition} then represent the coherence between the
electron and hole states $\varphi^{(\text{e})}_{l,i}$ and $\varphi^{(\text{h})}_{l',j}$
encoded in $g^{(\text{eh})}_{ij}(l-l')$. It can only be non-zero when the
electron and hole emission probabilities $p^{(\text{e})}_{i}$ and $p^{(\text{h})}_{j}$ are different
from $0$ or $1$. It then expresses the existence of a quantum superposition between
the unperturbed ground state and the creation of the electron/hole pair
built from the single-particle
states $\varphi^{(\text{e})}_{i}$ and $\varphi^{(\text{h})}_{j}$. The coefficients
$g^{(\text{eh})}_{ij}(l-l')$ encode the modulus and phase of such a quantum
superposition. In this description, the ideal emission of a quantized number
of $N_\text{e}$ electrons and $N_\text{h}$ holes is characterized by
$g^{(\text{ee})}_i(l-l')=\delta_{l,l'}$ and $g^{(\text{hh})}_i(l-l')=\delta_{l,l'}$
implying that
$g^{(\text{eh})}_{ij}(l-l')=0$.

This formalism serves as the theoretical background for the extraction of the
electron and hole wavefunctions from experimental measurements. Using two-particle
interferences, we proceed to the measurement of the electronic coherence
$\Delta_0 W$ and $\Delta_0 \mathcal{G}$ for arbitrary electrical currents. We then
implement an algorithm (see Methods) which identifies the emitted wavefunctions
$\varphi^{(\text{e})}_{i}$ and $\varphi^{(\text{h})}_{j}$ from the diagonalization of
$\Delta_0 \mathcal{G}$ in the subspace of electron and hole states and recasts it
in the form given by Eq.~\eqref{eq:final-decomposition}. This set of data
describes completely the single particle content of the electronic current and
quantifies how far it deviates from the ideal emission regime.\\

\begin{figure}[h!]
	 \includegraphics[width=\linewidth]{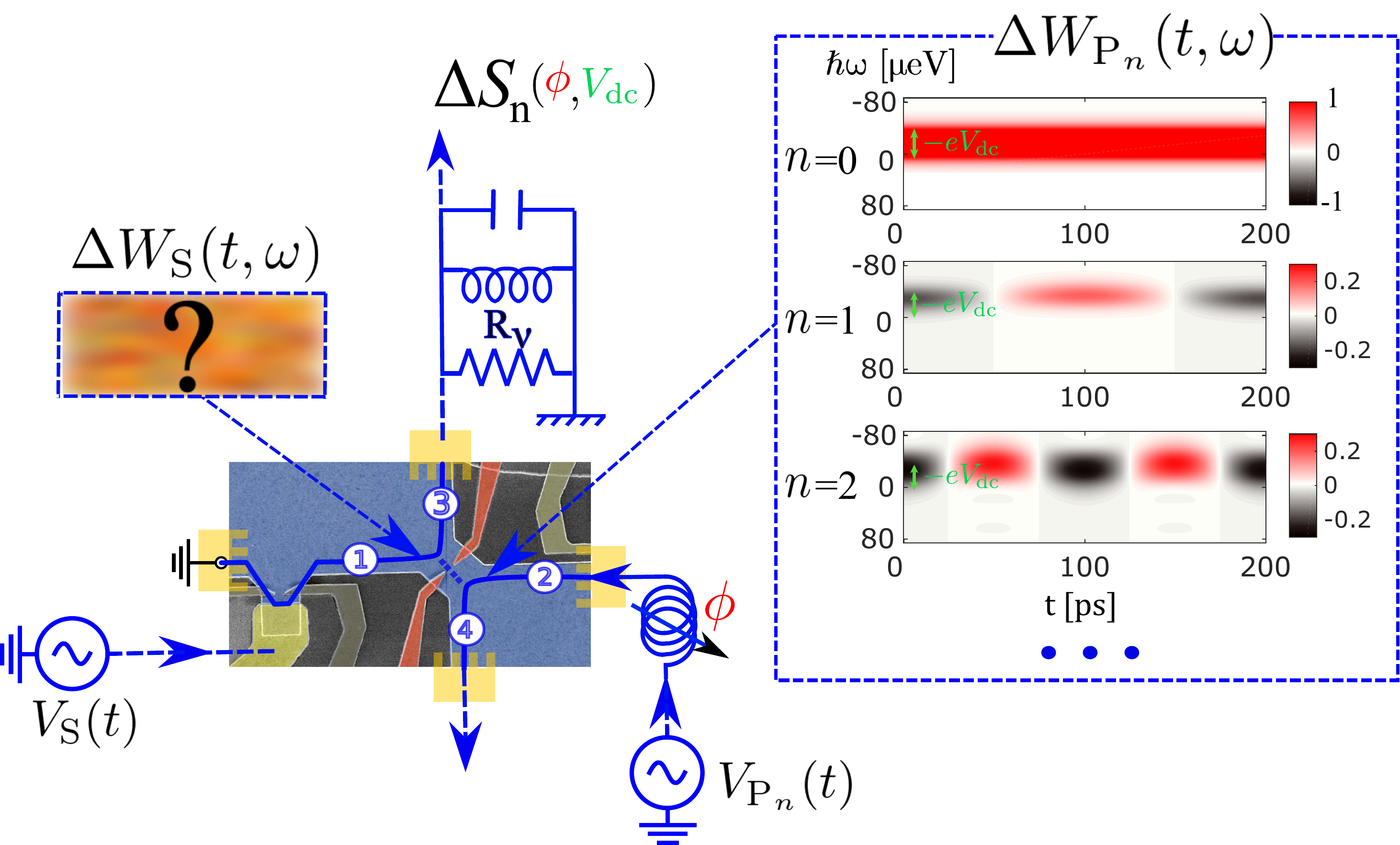}
	\caption{
Sketch of the experimental setup. The two dimensional electron gas is represented
in blue color and the edge channels as blue lines. A quantum point contact (red color)
is used as an electronic beam-splitter. A time dependent voltage $V_{\text{S}}(t)$ is applied to a
mesoscopic capacitor (gate in gold color capacitively coupled to the edge channel) placed at
input 1 of the beam-splitter and generates the unknown Wigner distribution $\Delta
W_{\text{S}}(t,\omega)$.  The probe signal $V_{\text{P}_n}(t)$, a low amplitude
sinusoidal drive at frequency $nf$ is generated at input 2.  The corresponding
probe Wigner distributions $\Delta W_{\text{P}_n}(t,\omega)$ are plotted for
$n=0$ to $n=2$ (the frequency is $f=\SI{5}{\giga\hertz}$ and the temperature
$T_{\text{el}}=\SI{80}{\milli\kelvin}$).  The current noise at the output of the
splitter is converted to a voltage noise on the quantized resistance
$R_{\nu}=h/(\nu \text{e}^2)$.  $R_{\nu}$ is connected to an LC tank circuit used to
shift the measurement frequency at the resonance $f_0=\SI{1.45}{\mega\hertz}$.
}
\label{fig2}
\end{figure}

{\bf Experimental setup and protocol.}
The experiment is performed in a high mobility GaAs/AlGaAs two-dimensional
electron gas placed in a strong perpendicular magnetic field so as to reach the
quantum Hall regime at filling factors $\nu=2$ or $\nu=3$ where charge
propagates along one-dimensional chiral edge channels.  We focus on the
propagation on the outer edge channel which realizes a ballistic spin-polarized
one-dimensional conductor. The electronic source is a metallic gate
capacitively coupled to the edge channels, allowing us to shape any charge
distribution \cite{Misiorny2018} by applying the proper time-dependent voltage
to the gate.  The resulting Wigner distribution $W_{\text{S}}(t,\omega)$ can be
reconstructed \cite{Tomo2011} by measuring two-electron interferences
\cite{Liu1998,Neder2007} using an electronic Hong-Ou-Mandel\cite{HOM}
interferometer \cite{Ol'khovskaya2008,Jonckheere2012,Bocquillon2013}.  As shown on
Fig.\ref{fig2}, the interferometer consists of a quantum point contact used as
an electronic beam-splitter partitioning the excitations propagating from inputs
1 and 2 with transmission probability $\mathcal{T}$.  Input 1 is connected to
the source whereas input 2 is connected to a voltage driven ohmic contact that
will generate a set of known reference states, called probe states, of Wigner
distribution $W_{\text{P}_n}$ for $n\in\mathbb{N}$.  For each probe state, we
measure the excess noise $\Delta S_n$ at output 3 between the source being
switched on and off\cite{Marguerite2017}:

\begin{equation}
\Delta S_n = 2 \text{e}^2 \mathcal{T}(1-\mathcal{T}) \int
 \frac{\md\omega}{2\pi}\left[ \overline{\Delta W_{\text{S}}(t,\omega)}^t
 \big(1-2f_{\text{eq}}(\omega)\big) \right.
	-\left.  2\overline{\Delta
 W_{\text{S}}(t,\omega)\Delta W_{\text{P}_n}(t,\omega)}^t \right]
	\label{SHOM}
\end{equation}

where $\overline{\cdots}^t$ denotes the average over time $t$, and
$\Delta W_{\text{S}/\text{P}_{n}}$ are respectively the source and probe excess Wigner
distribution with respect to the Fermi-Dirac distribution
$f_{\text{eq}}(\omega)$ at temperature $T_{\text{el}}\neq 0$:
$W_{\text{S}/\text{P}_{n}}(t,\omega)=f_{\text{eq}}(\omega) + \Delta
W_{\text{S}/\text{P}_{n}}(t,\omega)$.
The first term in Eq.~\eqref{SHOM}
represents the
classical random partition noise of the source.  It is reduced by the second
term in Eq.~\eqref{SHOM} which represents the antibunching between indistinguishable source and
probe excitations colliding on the splitter.  Their degree of
indistinguishability is given by the overlap between $\Delta W_{\text{S}}$ and
$\Delta W_{\text{P}_{n}}$.  By properly choosing the set of probe states,
Eq.~\eqref{SHOM} allows for the reconstruction of any unknown Wigner distribution
\cite{Tomo2011,DegioWigner2013}.

A convenient set of probe states can be used to reconstruct each harmonic of the
Fourier expansion of the excess source Wigner distribution:

\begin{equation}
\label{Wn}
\Delta W_{\text{S}}(t,\omega) = \sum_{n\in\mathbb{Z}} \Delta
W_{\text{S},n}(\omega)\,\me^{2 \pi \mi n f t},
\end{equation}

where
$f=1/T$ denotes the driving frequency. The $n=0$ harmonic represents the source excess electronic
distribution function $\Delta f(\omega)$. All the time dependance of $\Delta
W_{\text{S}}(t,\omega)$ is encoded in the $n \neq 0$ harmonics.  To select the
contribution from the $n^{\text{th}}$ harmonic in Eq.~\eqref{SHOM}, we apply on the
probe input a small ac signal at frequency $nf$ on top of a dc
bias\cite{Tomo2011}: $V_{\text{P}_n}(t)=V_{\text{dc}}+V_{\text{P}_n}\cos(2 \pi n f t +
\phi)$. The resulting Wigner distribution $W_{\text{P}_n}$ (plotted on
Fig.~\ref{fig2}) evolves periodically in time at frequency $n f$.  By measuring
the output noise $\Delta S_n$ as a function of $\phi$ and $V_{\text{dc}}$ (see
Methods), the real and imaginary parts of $\Delta W_{\text{S},n}(\omega)$ can be
extracted.\\

\begin{figure}[h!]
\includegraphics[width=\linewidth]{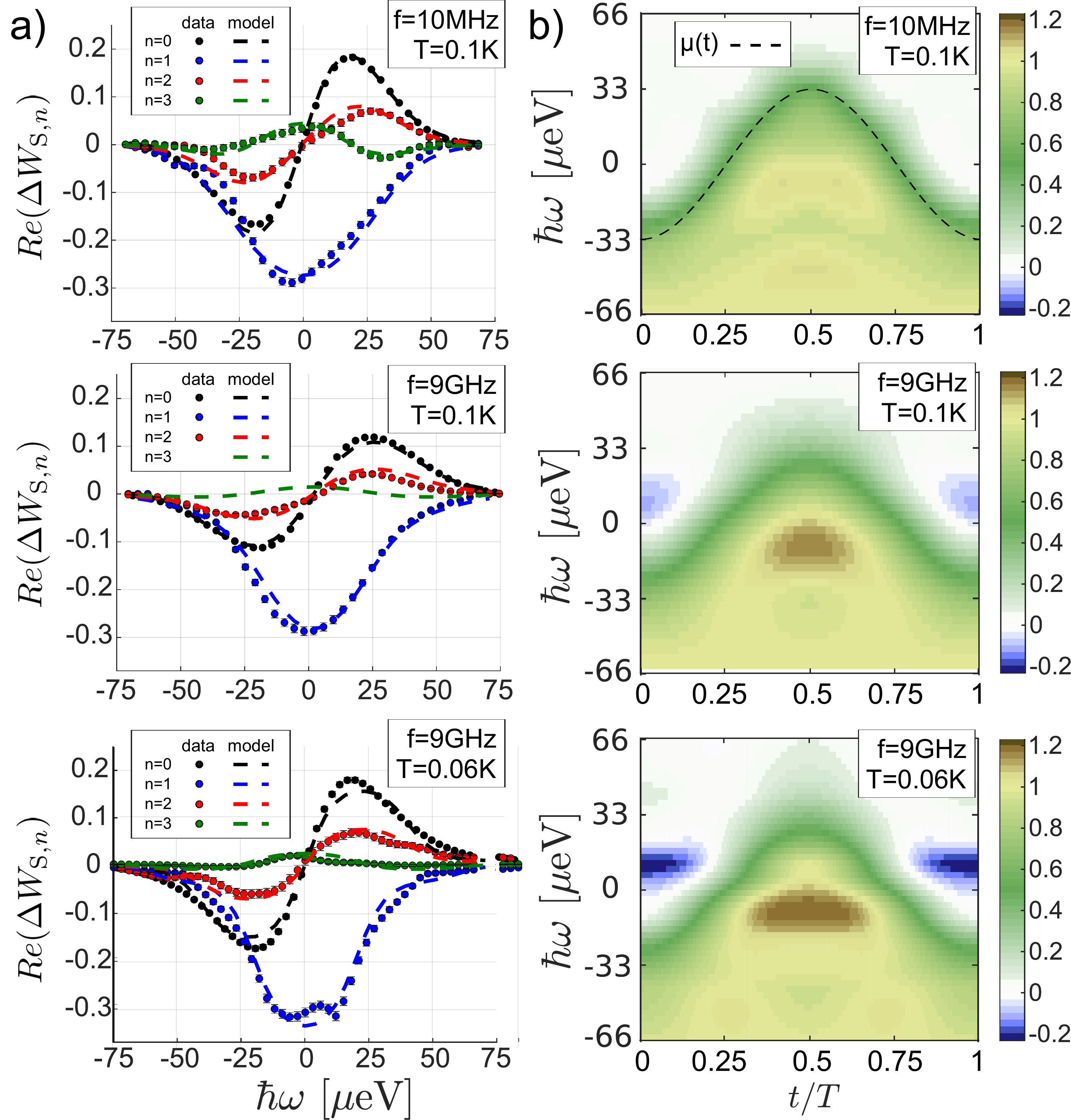}
	\caption{Wigner distribution of sinusoidal drives. \textbf{a} Measured
$\Delta W_{\text{S},n}(\omega)$ for $n=0$ to $n=3$ for sinusoidal drives at
frequency $f=\SI{10}{\MHz}$ and $f=\SI{9}{\GHz}$.  The observed parity: $\Delta
W_{\text{S},n}(-\omega)=(-1)^{n+1} \Delta W_{\text{S},n}(\omega)$, directly stems
from the electron/hole symmetry of the sinusoidal drive.  Error bars are defined
as standard error of the mean. \textbf{b}
Time-energy representation of the Wigner distribution $W_{\text{S}}(t,\omega)$.
}
\label{fig3}
\end{figure}

{\bf Electronic Wigner distribution of sinusoidal
drives.}
We first apply our quantum current analyzer to sinusoidal
drives, $V_\text{S}(t)=V_{\text{S}}\cos{(2 \pi f t)}$ at various frequencies $f$.  Figure
~\ref{fig3}a presents the measurements of the $n=0,1,2,3$ harmonics
of $\Re{(\Delta W_{\text{S},n})}$ ($\Im{(\Delta W_{\text{S},n})}=0$) of three
sinusoidal drives of similar amplitudes ($V_\text{S}\approx\SI{32}{\micro\volt }$).  We
first focus on the effect of frequency by comparing $\Delta W_{\text{S},n}$ for
$f=\SI{10}{\mega\hertz}$ and $f=\SI{9}{\giga\hertz}$ at
$T_{\text{el}}=\SI{100}{\milli \kelvin}$.  The $n=0$, $2$ and $3$ harmonics are
lower for $f=\SI{9}{\giga\hertz}$ compared to $f=\SI{10}{\mega\hertz}$ ($\Delta
W_{\text{S},n=3}$ even falls below our experimental resolution for
$f=\SI{9}{\giga\hertz}$).

Indeed, in a photo-assisted description of electronic
transport\cite{Reydellet2003,Rychkov2005}, the $n \neq 1$ harmonics are related
to multiphoton absorption/emission processes, whose strength increases with the
ratio $eV_{\text{S}}/h f$ which equals $800$ for $f=\SI{10}{\mega\hertz}$ compared to
$0.8$ for $f=\SI{9}{\giga\hertz}$.  We then turn to the effect of temperature by
comparing $\Delta W_{\text{S},n}$ for the two drives at $f=\SI{9}{\giga\hertz}$
but different temperatures.  Decreasing the temperature from
$T_{\text{el}}=\SI{100}{\milli \kelvin}$ to $T_{\text{el}}=\SI{60}{\milli
\kelvin}$ leads to a narrowing of all the harmonics and to an increase of their
amplitude.  For the three drives, the agreement between the data and theoretical
predictions (dashed lines) is excellent, showing the robustness of our
reconstruction procedure.  After measuring all relevant $\Delta W_{\text{S},n}$,
we can combine them in Eq.~\eqref{Wn} to reconstruct $W_{\text{S}}(t,\omega)$.

The Wigner distributions are represented on Fig.~\ref{fig3}b.
Within experimental accuracy, the $f=\SI{10}{\mega\hertz}$ case follows an
equilibrium distribution function, $W_{\text{S}}(t,\omega) = f_{\text{eq},
\mu(t)}(\omega)$, with a time varying chemical potential following the ac drive:
$\mu(t)=-eV_{\text{S}}\cos{(2\pi ft)}$.
This is expected as the
$f=\SI{10}{\mega\hertz}$ case corresponds to a quasi-classical current ($hf \ll
k_\text{B} T_{\text{el}}$) characterized by bounded values of the Wigner distribution,
$0 \leq W(t,\omega) \leq 1$, such that $W(t,\omega)$ can be
interpreted as a time-dependent electronic distribution function and viewed as
an adiabatic evolution of the stationary (dc) case\cite{DegioWigner2013}.
In
contrast, $hf \geq k_\text{B} T_{\text{el}}$ corresponds to the quantum case where the
Wigner distribution can take negative or above one values.  This is what we
observe for the $f=\SI{9}{\giga\hertz}$ drives, with a strong emphasis of these
quantum features at the lowest temperature $T_{\text{el}}=\SI{60}{\milli
\kelvin}$.  Consequently, in the quantum regime, single-particle properties are
no longer described in terms of a time varying electronic distribution function.
This is the case where $ W(t,\omega)$ can be used to extract electron and
hole wavefunctions.\\

\begin{figure}[h!]
	\includegraphics[width=1
\columnwidth,keepaspectratio]{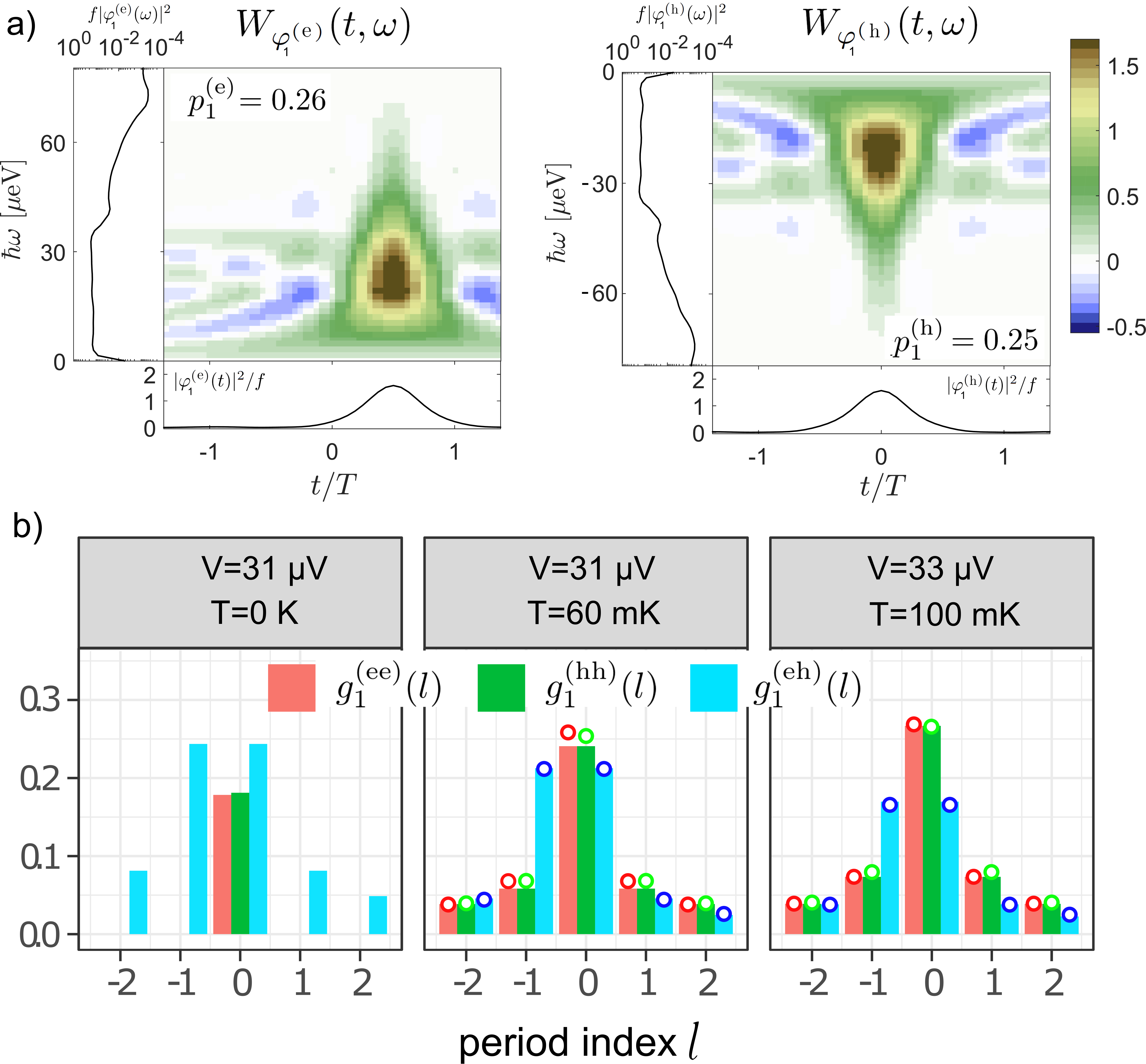}
	\caption{\label{fig4} Electron and hole wavefunctions generated by low amplitude sinusoidal drives.
\textbf{a}
Wigner distribution functions $W_{\varphi^{( \text{e/h})}_1}(t,\omega)=\int d\tau
\varphi^{(\text{e/h})}_1(t+\frac{\tau}{2})\varphi^{(\text{e/h}) *}_1(t-\frac{\tau}{2})
\,\me^{\mi\omega \tau}$ for the dominant electronic $\varphi^{(\text{e})}_1$ and hole
$\varphi^{(\text{h})}_1$ wavefunctions for $f=9~\mathrm{GHz}$ and
$T_{\text{el}}=\SI{60}{\milli\kelvin}$ ($\varphi^{(\text{e/h})}_1$ obtained at
$T_{\text{el}}=\SI{100}{\milli\kelvin}$ are almost identical).  The panels in the margins of
the color plots represent the time $|\varphi^{(\text{e/h})}_1(t)|^2/f$ and energy
$f|\varphi^{(\text{e/h})}_1(\omega)|^2$ distributions obtained by integrating
$W_{\varphi^{(\text{e/h})}_1}(t,\omega)$ over $\omega$ and $t$.  \textbf{b} Moduli
of the interperiod coherence $|g^{(\text{ee})}(l)|$, $|g^{(\text{hh})}(l)|$ and $|g^{(\text{eh})}(l)|$
(colored bars correspond to numerical calculations and colored dots to the
experimental data).  }
\end{figure}

{\bf Electron/hole wavefunctions generated by sinusoidal drives.}
The second step of our analyzer extracts individual electronic wavepackets from
the reconstructed Wigner distribution by implementing an algorithm (see Methods)
which recasts our measurements in the form of Eq.~\eqref{eq:final-decomposition}.
Figure \ref{fig4} presents the result of this analysis on the experimental data
obtained for the $f=\SI{9}{\giga\hertz}$ sinusoidal drives.
As the probability
to emit more than one electron/hole is very small, the analysis can be limited
to one electron $\varphi^{(\text{e})}_1$ and one hole $\varphi^{(\text{h})}_1$ wavefunction
($p^{(\alpha)}_{i>1} \approx 10^{-3} \ll 1$).  They are plotted in the Wigner representation in
the case $T_{\text{el}}=\SI{60}{\milli \kelvin}$ on
Fig.~\ref{fig4}a.  The hole is shifted by half a period with respect to the
electron and its energy distribution $|\varphi^{(\text{h})}_1(\omega)|^2$ mirrors
that of the electron's at positive energy.  As a figure of merit of the
procedure, we evaluate the state fidelity defined as the overlap between
electron and hole wavefunctions extracted from the experimental data and the
electron and hole wavefunctions extracted from numerical computations of the
Wigner distribution using Floquet scattering theory (see Supplementary Note 1).
The results are in excellent agreement
with a fidelity greater than $0.99$ for all the extracted wavefunctions,
demonstrating the accuracy of the state reconstruction.

Fig.~\ref{fig4}b depicts the moduli of the inter-period
coherences between emitted wavepackets for different temperatures (bars
represent numerical simulations, circles represent data). When the temperature
increases, the occupation probabilities which, in the present case, are very close due to
electron/hole symmetry ($p_1^{(\text{e})} \approx p_1^{(\text{h})}$) increase
from $0.17$ (numerical calculation at $T_{\text{el}}=\SI{0}{\kelvin}$) to $0.25$
($\SI{60}{\milli\kelvin}$) and $0.27$ ($\SI{100}{\milli\kelvin}$).  Coherences
between different periods ($g_1^{(\text{ee})}(l\neq 0)$ and $g_1^{(\text{hh})}(l\neq 0)$ ) also
appear and extend over the thermal coherence time ($h/k_\text{B}T_{\text{el}}\simeq
\SI{0.5}{\nano\second}$ at $\SI{100}{\milli\kelvin}$).  This reflects that at
finite temperature, the electron and hole states $\varphi_1^{(\text{e/h})}$ have a finite
probability to be occupied by thermal excitations.  As the probability to emit
the electron and hole differ from $1$, we also observe non-zero electron/hole
coherence: $g_{11}^{(\text{eh})}(l-l')\neq 0$.  Interestingly, these terms are suppressed by
thermal fluctuations, reflecting the transition from a pure quantum
state at
$T_{\text{el}}=\SI{0}{\kelvin}$ to a statistical mixture at higher
temperature.  At $T_{\text{el}}=\SI{0}{\kelvin}$,
a single process occurs: the generation of
the quantum superposition between the unperturbed ground state (with probability
$1-p_1^{(\text{e})}$) and the creation of the electron/hole pair (with probability
$p_1^{(\text{e})}$).
Thermal fluctuations allow two additional processes: only the electron state, or
only the hole state, can be generated. The resulting state at finite
temperature is a statistical mixture between these three processes.  Our
algorithm enables a quantitative description by computing a purity indicator,
$\mathbb{P}$, from the extracted inter-period coherences (see Methods). It
quantifies the weight of coherent electron/hole processes with respect to all
emitted excitations.  By construction $\mathbb{P}=1$ at zero temperature and,
from our experimental data, decreases to $0.71$ at
$T_{\mathrm{el}}=\SI{60}{\milli\kelvin}$ and to $0.58$ at
$T_{\mathrm{el}}=\SI{100}{\milli\kelvin}$.  Numerical evaluation of the same
quantity calculated using Floquet scattering theory (see Supplementary Note 1) give $0.999$ at zero temperature, $0.725$ at
$\SI{60}{\milli\kelvin}$ and $0.588$ at $\SI{100}{\milli\kelvin}$ in very good
agreement with the experimental data.\\

{\bf Single electron Lorentzian pulse ($\textbf{q=-\text{e}}$)}.
We now turn to the analysis of a
current generated by periodic
Lorentzian voltage pulses $V_\text{S}(t)=\sum_l -\frac{V_0}{1+(t-lT)^2/\tau^2}$ with
$\tau=\SI{42}{\pico\second}$, $f=\SI{4}{\giga\hertz}$ and $V_0$ chosen such that
each pulse carries exactly a single-electron charge:
$\frac{\text{e}^2}{h}\int_0^{T}V_\text{S}(u)du=-\text{e}$. The Lorentzian pulses are generated by calibrating
 the amplitude and phase of each harmonic of the current at the location of the beam-splitter (see Methods).
 This ensures that phase and amplitude shifts caused by Coulomb interaction effects\cite{Grenier2013,Cabart2018} during the propagation along the edge channels can be absorbed in the calibration process. This procedure allows us
 to demonstrate the proof of principle of our quantum current tomography without having to consider Coulomb interaction
 effects.

\begin{figure}[h!]
	\includegraphics[width=1
\columnwidth,keepaspectratio]{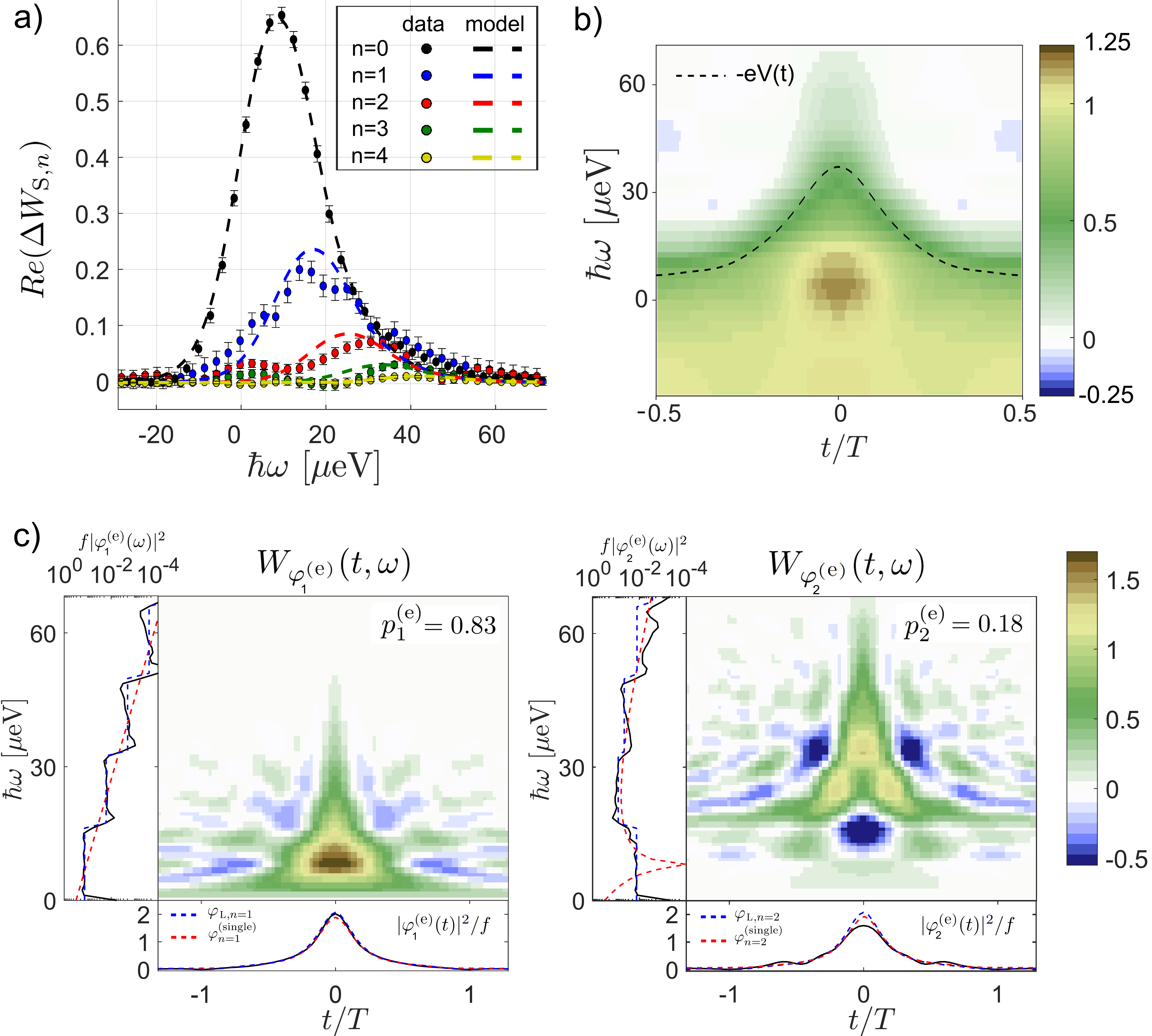}
	\caption{\label{fig5} Electron wavefunction generated by a Lorentzian pulse, $q=-\text{e}$. \textbf{a} Measured
$\Delta W_{\text{S},n}(\omega)$ ($n=0$ to $n=4$) for the single electron
Lorentzian pulse at $f=\SI{4}{\giga\hertz}$, $\tau=42$ ps and
$T_\mathrm{el}=\SI{50}{\milli\kelvin}$. Error bars are defined as standard error of the mean.
\textbf{b} $W_{\text{S}}(t,\omega)$, the dashed
line represents the voltage pulse $V(t)=h/\text{e}^2 I(t)$ where $I(t)$ is obtained by
integrating $W_{\text{S}}(t,\omega)$ on energy $\omega$.
\textbf{c} Wigner
representation of $\varphi^{(\text{e})}_1(t)$ (left) and $\varphi^{(\text{e})}_2(t)$ (right).
The panels in the margins represent the time $|\varphi^{(\text{e})}_i(t)|^2$ and energy
$|\varphi^{(\text{e})}_i(\omega)|^2$ distributions obtained by integrating
$W_{\varphi^{(\text{e})}_i}(t,\omega)$ over $\omega$ and $t$.
$|\varphi^{(\text{e})}_1(\omega)|^2$ is represented in log scale for better comparison
	with theoretical predictions with single shot
	$\varphi^{(\text{single})}_{n}$ (red
dashed line) and periodic $\varphi_{\text{L}, n}$ (blue dashed line) Lorentzian
wavepackets (with $\tau=42$ ps).  As it can be seen from the very good agreement
with the blue dashed line, the steplike behaviour of $|
\varphi^{(\text{e})}_1(\omega)|^2$ comes from the periodicity of the drive.  }
\end{figure}

  The measured $n=0$ to $n=4$ harmonics of
$\Delta W_{\text{S}}(t,\omega)$ are plotted on Fig.~\ref{fig5}a.  They are
located on the positive energy side with maxima shifted by $n h f/2\text{e}$ for
increasing $n$ and, indeed, take very small value
for $|\omega|\leq
-n\pi f$ showing that almost no hole excitation is emitted.
Their energy width is imposed by the temporal width of the pulse
$\tau$ and their amplitude reproduces the decrease of the harmonics of $I(t)$
(obtained by integration of $\Delta W_{\text{S},n}(\omega)$).  The overall
agreement with theoretical predictions (dashed lines) is good (no fitting
parameter).  The resulting Wigner distribution $W_{S}(t,\omega)$ is
plotted on Fig.~\ref{fig5}b. Strong non-classical features
($W_{\text{S}}(t,\omega) \approx 1.2$) are observed at the location of the
electron excitation in the $(t,\omega)$ plane.

The wavefunctions extracted from
our analysis are plotted on Fig.~\ref{fig5}c. Firstly, as expected for a single
electron Lorentzian pulse, this analysis quantitatively confirms that
almost no hole excitation is emitted:
$p^{(\text{h})}=0.03\pm 0.01$.  Secondly, contrary to the previous case (low amplitude
sine drive), two electronic wavefunctions $\varphi^{(\text{e})}_1$ and
$\varphi^{(\text{e})}_2$ contribute with emission probabilities $p^{(\text{e})}_1=0.83\pm
0.01$ and $p^{(\text{e})}_2=0.18 \pm 0.01$.  This means that finite temperature
(depending on the ratio $k_\text{B} T_{\text{el}}/hf$, see Supplementary Note 2) leads to the generation of a
statistical mixture\cite{Moskalets2017} (of purity $\mathbb{P}=0.75$) between
$\varphi^{(\text{e})}_1$ and $\varphi^{(\text{e})}_2$ with weights $p^{(\text{e})}_1$ and $p^{(\text{e})}_2$
($p^{(\text{e})}_1+ p^{(\text{e})}_2=1.01 \pm 0.01$). Note that such an emission of a statistical mixture
between different single electron wavefunctions could not be captured within the single body limit
considered in the previous analysis of Jullien \textit{et al.}\cite{Jullien2014} which underlines the need for
developing the general approach we demonstrate here.

The wavefunctions
$\varphi^{(\text{e})}_1$ and $\varphi^{(\text{e})}_2$ can be compared with the expected
ones generated by Lorentzian
voltage pulses at zero temperature. A single Lorentzian pulse of temporal
width $\tau$ and carrying an integer number $N_\text{e}$ of electrons is the
Slater determinant built over the $1\leq n\leq N_\text{e}$ electronic
wavefunctions\cite{Keeling06,Grenier2013}
$\varphi^{(\text{single})}_{n}(\omega)= \frac{1}{\sqrt{\mathcal{N}}} \Theta(\omega)
\exp(-\omega\tau) \mathcal{L}_{n-1}(2\omega \tau)$, where $\mathcal{N}$ is a
normalization constant, $\mathcal{L}_{n}$ is
the $n^{\text{th}}$ Laguerre polynomial. The two first ones
$\varphi^{(\text{single})}_{n=1}(\omega)$ and
$\varphi^{(\mathrm{single})}_{n=2}(\omega)$, plotted in red dashed lines on
Fig.~\ref{fig5}c, are very similar to $\varphi^{(\text{e})}_1$ and
$\varphi^{(\text{e})}_2$. However, they do not reproduce the discretization of the
energy distribution in units of $hf$. This discretization is related to the
periodicity of the single electron emission, which is not captured by the
expression of $\varphi^{(\text{single})}_{n}$. The $n=1$ wavefunction of the
periodic train of Lorentzian pulses has been
shown\cite{Moskalets2015,These:Roussel} to be
given by $\varphi_{\text{L},n=1}(\omega)= \frac{1}{\sqrt{\mathcal{N}}} \Theta(\omega)
\exp(-\omega_1 \tau)$ where $\omega_1=2 \pi f \left \lfloor{\frac{\omega}{2\pi
f}} \right \rfloor$ has quantized steps $2\pi f$ related to the pulse
periodicity. Here, we generalize this
expression to the $n=2$ and $n=3$ wavefunctions by using the following ansatz:
$\varphi_{\text{L},n}(\omega)= \frac{1}{\sqrt{\mathcal{N}}} \Theta(\omega)
\exp(-\omega_n\tau) \mathcal{L}_{n-1}(2\omega_n \tau)$ where $\omega_n=2 \pi f
\left(x_n+ \left\lfloor{\frac{\omega}{2\pi f}}\right\rfloor \right)$. We take $x_1=0$
following Refs.~\onlinecite{Moskalets2015,These:Roussel}
and then numerically deduce $x_2=0.33$
and $x_3=0.24$ from the constraint that the wavefunctions should be orthogonal:
$\langle\varphi_{\text{L},n}|\varphi_{\text{L},n'}\rangle=\delta_{n,n'}$.
Comparing the wavefunctions $\varphi^{(\text{e})}_1$ and
$\varphi^{(\text{e})}_2$ extracted from our experimental data to these theoretical
predictions, we observe that $\varphi^{(\text{e})}_1$ is very close to the expected
wavefunction $\varphi_{\text{L},n=1}$ (blue dashed line on Fig.~\ref{fig5}c) with an
overlap of $0.98$.  Interestingly, $\varphi^{(\text{e})}_2$ which is emitted at higher
energy strongly resembles $\varphi_{\text{L},n=2}$ with an overlap of $0.93$ leading to
this simple interpretation of temperature effects: finite temperature leads to
the emission of a statistical mixture between the expected $n=1$ Lorentzian
wavefunction and the wavefunctions corresponding to higher excitations numbers
$n>1$.  Importantly, our observations are not related to imperfections of the
emission drive or to errors of our extraction method but only to thermal
effects. This probabilistic description of the electron state stems from the
non-zero entropy of the finite temperature ground state which reveals the
statistical (non-quantum) fluctuations of the ground state. This interpretation
is confirmed by numerical calculations. Applying our method on perfect periodic
Lorentzian pulses calculated using Floquet scattering theory (see Supplementary Note 2), we
recover that for $T_{\mathrm{el}}=\SI{0}{\kelvin}$, $p^{(\text{e})}_1=1$ and $p^{(\text{e})}_2=0$ (pure
state) but for $T_{\mathrm{el}}=$\SI{50}{\milli\kelvin}, $p^{(\text{e})}_1=0.84$ and
$p^{(\text{e})}_2=0.18$, in very good agreement with our experimental results.\\

\begin{figure}[h!]
	\includegraphics[width=1
\columnwidth,keepaspectratio]{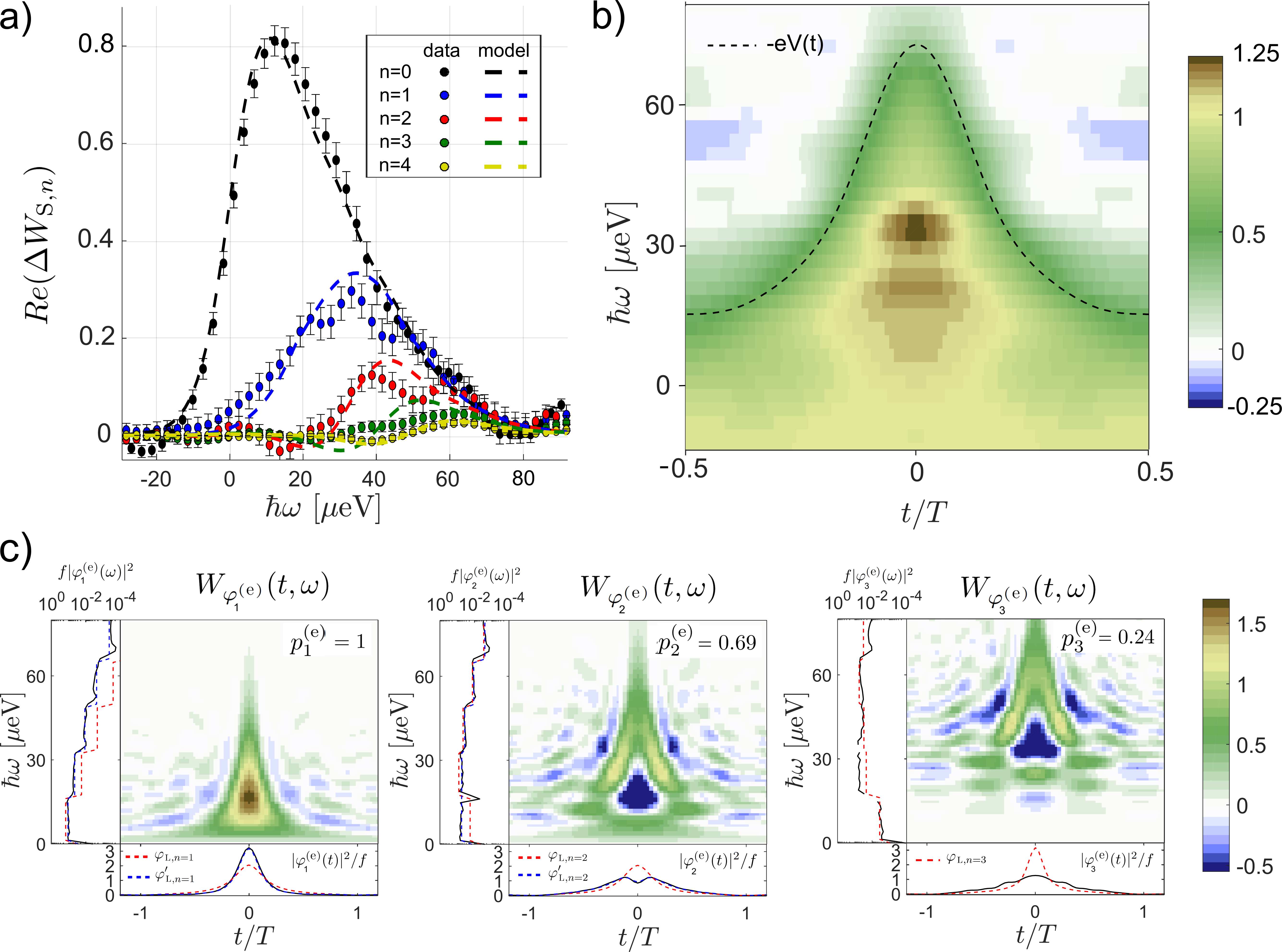}
	\caption{\label{fig6} Electron wavefunction generated by a Lorentzian pulse, $q=-2\text{e}$. \textbf{a} Measured
$\Delta W_{\text{S},n}(\omega)$ ($n=0$ to $n=4$) for the two electrons Lorentzian
pulse at $f=\SI{4}{\giga\hertz}$, $\tau=42$ ps and
$T_\mathrm{el}=\SI{50}{\milli\kelvin}$. Error bars are defined as standard error of the mean.
\textbf{b} $W_{\text{S}}(t,\omega)$, the dashed
line represents the voltage pulse $V(t)=h/\text{e}^2 I(t)$ where $I(t)$ is obtained by
integrating $W_{\text{S}}(t,\omega)$ on energy $\omega$.
\textbf{c} Wigner
representation of $\varphi^{(\text{e})}_1(t)$ (left) and $\varphi^{(\text{e})}_2(t)$ (middle)
and $\varphi^{(\text{e})}_3(t)$ (right).  The panels in the margins represent the time
$|\varphi^{(\text{e})}_i(t)|^2$ and energy $|\varphi^{(\text{e})}_i(\omega)|^2$ distributions
obtained by integration of $W_{\varphi^{(\text{e})}_i}(t,\omega)$ over $\omega$ and
$t$.  $|\varphi^{(\text{e})}_1(\omega)|^2$ is represented in log scale.  The red dashed
line represent the theoretical predictions for the $n=1$ to $n=3$ wavefunctions
of periodic trains of Lorentzian pulses $\varphi_{\text{L}, n}$.  The blue dashed lines
represent the theoretical predictions for $\varphi'_{\text{L},n}$ obtained from linear
combinations of the $\varphi_{\text{L}, n}$.}
\end{figure}

{\bf Two electrons Lorentzian pulse ($\textbf{q=-2\text{e}}$).}
Finally, we analyze periodic trains of Lorentzian pulses carrying the
charge of two
electronic excitations: $V_\text{S}(t)=\sum_l -\frac{2 V_0}{1+(t-lT)^2/\tau^2}$.  Our
experimental results are plotted on Fig.~\ref{fig6}.  As in the single electron
case, thermal effects lead to the generation of a statistical mixture of
different wavefunctions, one more than the number of emitted charges. The
first wavefunction $\varphi^{(\text{e})}_1$ is emitted with unit probability
$p^{(\text{e})}_1=1$ but $\varphi^{(\text{e})}_2$ and $\varphi^{(\text{e})}_3$ are emitted with
probabilities smaller than one, $p^{(\text{e})}_2=0.69\pm0.02$ and $p^{(\text{e})}_3=0.24\pm 0.02$,
reflecting that the emitted state is a statistical mixture of purity
$\mathbb{P}=0.68$.

Interestingly, and contrary to the $q=-\text{e}$,
case, $\varphi^{(\text{e})}_1$ and $\varphi^{(\text{e})}_2$ do not correspond to the expected
Lorentzian wavefunctions $\varphi_{\text{L}, n=1}$ and $\varphi_{\text{L}, n=2}$ plotted in
red dashed line on Fig.~\ref{fig6}c.  This can be understood by discussing first
the zero temperature case.  At $T_{\mathrm{el}}=\SI{0}{\kelvin}$, the generated state is
predicted to be described by the Slater determinant formed from the
wavefunctions $\varphi_{\text{L}, n=1}$ and $\varphi_{\text{L}, n=2}$.  However, any choice of
basis obtained by a linear combination of $\varphi_{\text{L}, n=1}$ and $\varphi_{\text{L},
n=2}$ is equally valid to describe this Slater determinant (see Supplementary Note 3).  At
finite temperature, $T_{\mathrm{el}}\neq\SI{0}{\kelvin}$, this
ambiguity in the choice of the two
wavefunctions describing the electronic state is lifted as $\varphi^{(\text{e})}_1$ and
$\varphi^{(\text{e})}_2$ are no longer generated with the same
probability: $p^{(\text{e})}_1\neq p^{(\text{e})}_2(\neq 1)$.

Searching for the basis of states
which maximizes the overlap with $\varphi^{(\text{e})}_1$ and $\varphi^{(\text{e})}_2$, we
observe that finite temperature favors the emergence of two specific states
obtained by the following linear combination of $\varphi_{\text{L}, n=1}$ and
$\varphi_{\text{L}, n=2}$:
\begin{subequations}
\begin{align}
	\varphi'_{\text{L}, n=1} &=
\cos\left(\frac{\theta}{2}\right)\,\varphi_{\text{L}, n=1} +
	\sin\left(\frac{\theta}{2}\right)\,
\varphi_{\text{L}, n=2}\\
	\varphi'_{\text{L}, n=2} &= \sin\left(\frac{\theta}{2}\right) \;\varphi_{\text{L},
n=1}-\cos\left(\frac{\theta}{2}\right)\, \varphi_{\text{L}, n=2}
\end{align}
\end{subequations}
with $\theta\sim\frac{\pi}{2}\times 0.37$.
The single electron wavefunctions
$ \varphi'_{\text{L}, n=1} $ and
$\varphi'_{\text{L}, n=2} $, plotted in blue dashed lines on Fig.~\ref{fig6}c, have a
very strong overlap with $\varphi^{(\text{e})}_1$ and $\varphi^{(\text{e})}_2$ ($0.99$ and
$0.96$). As $p^{(\text{e})}_1=1$ and $p^{(\text{e})}_2=0.69$, it means that with probability
$0.69$ the two electron state described by the Slater determinant formed from
$\varphi'_{\text{L},1}$ and $\varphi'_{\text{L},2}$ is generated. This state is equivalent to
the expected Slater determinant formed from $\varphi_{\text{L}, 1} $ and
$\varphi_{\text{L},2} $. However, with probability $p^{(\text{e})}_2=0.24$
a different two electron
state is generated corresponding to the Slater determinant formed from
$\varphi'_{\text{L},1}$ and $\varphi^{(\text{e})}_3$.

The connection between $\varphi^{(\text{e})}_3$ and a theoretically
predicted wavefunction is less clear.
$|\varphi^{(\text{e})}_3(\omega)|^2$ resembles $ |\varphi_{\text{L}, n=3}(\omega)|^2$, the
energy distribution of the $n=3$ Lorentzian wavefunction (blue dashed
lines on Fig.~\ref{fig6}c).  However the time distributions are different which
is reflected by the relatively small overlap of $0.86$ between the two states.

As for the $q=-\text{e}$ case, we can check using numerical computations that the
generation of a statistical mixture between two different Slater
determinants is caused by the finite temperature. The simulations presented in
Supplementary Note 3 confirm that the probability to generate the Slater determinant formed
from $\varphi_{\text{L},1}$ and $\varphi_{\text{L},2}$ decreases from 1 at zero temperature to
$0.79$ at $T_{\mathrm{el}}=\SI{50}{\milli\kelvin}$ while the probability to
generate the Slater determinant formed
from $\varphi'_{\text{L},1}$ and $\varphi^{(\text{e})}_3$  increases from 0 at zero
temperature to $0.18$ at $T_{\mathrm{el}}=\SI{50}{\milli\kelvin}$ in reasonable
agreement with our observations.

\section*{Discussion}
We have demonstrated a quantum tomography protocol for arbitrary electrical
currents.  Without any a priori knowledge on the electronic state, this protocol
extracts all the electron and hole wavefunctions and their emission
probabilities.

Our protocol is the tool of choice for characterizing single to few
electron sources by extracting all the $N_\text{e}$ single electron wavefunctions
generated at each period.  We have analyzed in this work the $N_\text{e}=1$ and $N_\text{e}=2$ Lorentzian
voltage pulses and have extracted the wavefunctions of each generated electronic excitation.
The generation of Lorentzian electronic wavepacket in a ballistic
one-dimensional channel which we demonstrate here is an important milestone for the
development of time resolved quantum electronics. Numerous recent proposals
suggest the use of time resolved charge or energy currents\cite{Dashti2019}
carried by integer charge Lorentzian voltage pulses to probe the timescales
of quantum coherent conductors\cite{Burset2018} or to dynamically control
their interference pattern\cite{Gaury2014}.

The reconstruction of the quantum state of single electronic excitations is a
currently active research field as illustrated by the
very recent achievement of the state tomography of high energy electrons
propagating along quantum Hall edge channels\cite{Fletcher2019,Locane2019}.
These recent works highlight the importance of characterizing the purity of the
emitted states. Importantly, our protocol fully captures the differences between
pure states and statistical mixtures and provides a quantitative evaluation of
the purity. This ability to quantify the purity of quantum states generated by
electronic sources is crucial for future applications of quantum
electronics.  More specifically, we find that for single charge
Lorentzian pulses, which are predicted to generate a pure single electron
wavefunction at zero temperature, poisoning by thermal excitations results in
the emission of a mixture (with purity $\mathbb{P}=0.75$) of two different
states which correspond to the $n=1$ and $n=2$ Lorentzian wavepackets
$\varphi_{\text{L},n}$. For two electron Lorentzian pulses, we show that thermal effects
lead to the generation of a mixture between the zero temperature Slater determinant
formed from $\varphi_{\text{L},1}$ and $\varphi_{\text{L},2}$ and an undesired two electron state
which we fully characterize.

The generation and characterization of multi-electron states in
quantum conductors also opens the way to the study of correlations
and interactions between a controlled number of excitations emitted on demand in
the circuit, with applications to the controlled generation of entangled
electron or electron/hole\cite{Hofer:2016-2} pairs. In this context, this
protocol can also be applied to identify single particle wavefunctions generated
in interacting conductors\cite{Cabart2018} and supplemented by other
measurements\cite{Thibierge2016}, to quantify the importance of
interaction-induced quantum correlations.

Finally, it can establish a bridge
between electron and microwave quantum optics
\cite{Grimsmo:2016-1,Virally:2016-1} by probing the electronic content of
microwave photons injected from a transmission line into a quantum conductor.

\section*{Methods}

\subsection{Sample and noise measurements}

The sample is a GaAs/AlGaAs two dimensional electron gas of charge density
$n_\text{s}=\SI{1.9e15}{\per\square\meter}$ and mobility
$\mu=\SI{2.4e6}{\per\square\centi\meter\per\volt\per\second}$.  It is placed in
a high magnetic field to reach the quantum Hall regime at filling factor $\nu=2$
($B=\SI{3.7}{\tesla}$) and $\nu=3$ ($B=\SI{2.6}{\tesla}$).  The measurements of the
Wigner distribution of sinusoidal drives have been performed at $\nu=2$. The measurements of
the Wigner distributions of Lorentzian pulses have been performed at $\nu=3$.  The current noise
at the output of the quantum point contact is converted to a voltage noise via
the quantum Hall edge channel resistance $R_{\nu}=h/\nu \text{e}^2$ between output 3
and the ohmic ground (see Fig.~\ref{fig2}).  In order to move the noise
measurement frequency in the $\si{\MHz}$ range, $R_{\nu}$ is connected to an LC tank
circuit of resonance frequency $f_0=\SI{1.45}{\mega\hertz}$.  The tank circuit
is followed by a pair of homemade cryogenic amplifiers followed by room
temperature amplifiers.  A vector signal analyzer measures the correlations
between the voltages at the output of the two amplification chains in a
$\SI{78}{\kilo\hertz}$ bandwidth centered on $f_0$.  The noise measurements are
calibrated by measuring both the thermal noise of the output resistance
$R_{\nu}$, and the partition noise of a d.c.  bias applied on input 2 of the
electronic beam-splitter.

\subsection{Generation of Lorentzian current pulses}

The single electron and two electrons periodic trains of Lorentzian current
pulses are generated by applying the ac part of the signal $V_{\text{ac}}(t)$ on the
mesoscopic capacitor placed at input $1$ of the beam-splitter, see
Fig.\ref{fig1}, and the dc part of the signal $V_{\text{dc}}$ on the ohmic contact
connected to input $1$ of the beam splitter such that $V_\text{S}(t)=V_{\text{ac}}(t)+V_{\text{dc}} =
\sum_l -\frac{V_0}{1+(t-lT)^2/\tau^2}$.  More precisely, $V_{\text{ac}}(t)$ is
generated harmonic by harmonic (from $n=1$ to $n=5$): $V_{\text{ac}}(t) =
\sum_{n=1}^{n=5} V_{\text{ac},n} \cos(n\Omega t + \phi_n)$.  The careful calibration
of the amplitude $V_{\text{ac},n}$ and phase $\phi_n$ of each harmonic is performed at
the location of the beam-splitter using low-frequency noise measurements.  The
calibration of each amplitude is performed by sending a single harmonic $
V_{\text{ac},n} \cos(2\pi n f t + \phi_n)$ towards the splitter and measuring the
low-frequency noise as a function of $V_{\text{ac},n}$.  The calibration of the
relative phases $\phi_n$ is more difficult and involves two-particle
interferences between two harmonics at two different frequencies.  As an
example, to calibrate the relative phase $\phi_n$ between the first and $n^{\text{th}}$
harmonic of the signal, we generate the voltage $V_\text{S}(t)=V_{\text{ac},1}\cos(2\pi f t)
+ V_{\text{ac},n}\cos(2 \pi n f  t + \phi_n)$ at input 1 of the splitter.  From
two-particle interference effect, the noise at the splitter output depends on
the relative phase $\phi_n$ between the two harmonics.  It is minimal (or
maximal depending on the harmonics considered) when the two harmonics are in
phase, allowing for an accurate calibration of the relative phase between the
different harmonics.  Note that for even harmonics, the two-particle
interferences between the two harmonics vanish at zero bias voltage such that a
small bias voltage of a few tens of microvolts needs to be added for their phase
calibration.  The fact that amplitude and phase of each harmonic is calibrated
at the splitter location is very important regarding the effect of Coulomb
interaction during the propagation of single electron excitations.  Previous
works have emphasized\cite{Wahl2014} the importance of these effects in quantum
Hall edge channels and their impact on the relaxation and
decoherence\cite{Marguerite2016} of electronic excitations or on the
fractionalization\cite{Freulon2015} of these excitations.  In our experimental
setup, single or two-electron Lorentzian wavepackets will eventually
fractionalize\cite{Grenier2013} due to Coulomb interaction effect.  However,
this fractionalization will occur after the beam-splitter where we performed the
tomography experiment.  Indeed, these single electron excitations result from
the generation of voltage pulses propagating along the edge channels.  For such
types of electronics sources, which are qualitatively different from quantum dot
emitters where electron emission involves the tunneling through a transmission
barrier, Coulomb interaction can be taken into account by a
renormalization\cite{Grenier2013} of the amplitude and phase of each harmonic of
the voltage pulse.  By calibrating these amplitudes and phases at the splitter
location, it means that we absorb the effect of Coulomb interaction by
accommodating the amplitudes and phases of the signal to reconstruct at the
splitter a Lorentzian pulse that is only limited by our calibration accuracy.
This allows us to neglect Coulomb interaction effects in this experiment
contrary to previous experiments\cite{Bocquillon2013,Marguerite2016} performed
with ac driven quantum dots.

\subsection{Reconstruction of $\Delta W_{\text{S},n}(\omega)$ from noise
measurements}

In this paper, we implement a reconstruction of the source Wigner distribution
$W_{\text{S}}(t,\omega)$ from the measurement of the current noise $\Delta S_{n}$
that does not rely on any assumption on the electronic state generated by the
source.

Firstly, the excess electronic distribution function $\Delta
W_{\text{S},n=0}(\omega)$ can be obtained via the derivative of the noise $\Delta
S_{n=0}$ with respect to the d.c.  bias
$\omega_{\text{dc}}=-eV_{\text{dc}}/\hbar$ applied on the probe port 2.
\begin{widetext}
	\begin{subequations}
\begin{align}
	\Delta S_{n=0} &= 2 \text{e}^2 \mathcal{T}(1-\mathcal{T}) \int
 \frac{\md\omega}{2\pi}\left[ \overline{\Delta W_{\text{S}}}^t
 \left(1-2f_{\text{eq}}\left(\omega - \omega_{\text{DC}}\right)\right) \right]
 \\
\widetilde{\Delta W}_{\text{S},0} &= -\frac{\pi } {2 \text{e}^2
\mathcal{T}(1-\mathcal{T})} \frac{\partial \Delta S_{n=0}}{\partial
\omega_{\text{DC}}} = \int d\omega \Delta W_{\text{S},0}\left(\omega\right)
\left(\frac{-\partial f_{\text{eq}}}{\partial \omega}\right)\left(\omega -
\omega_{\text{DC}}\right)
\label{n0}
\end{align}
	\end{subequations}
\end{widetext}
As shown by Eq.\eqref{n0}, the experimental signal $\widetilde{\Delta
W}_{\text{S},0}$ does not provide directly $\Delta W_{\text{S},0}$ but its
convolution with the thermally broadened function $\left(\frac{-\partial
f_{\text{eq}}}{\partial \omega}\right)$.  Knowing the electronic temperature,
one can reconstruct $\Delta W_{\text{S},0}(\omega)$ from the measurement of
$\widetilde{\Delta W}_{\text{S},0}$ using Bayesian deconvolution techniques
presented in the next section.

We then turn to the higher order terms, $\Delta W_{\text{S},n \neq 0}$, which
encode all the time dependence.  They are reconstructed from the measurement of
the output noise $\Delta S_{n,\phi}$ as a function of the d.c.  voltage
$V_{\text{DC}}$ and the phase $\phi$ of the a.c.  voltage $V_{\text{P}_n}(t)$ applied
on the probe port.  For a sinusoidal drive at frequency $nf$ of small amplitude
$V_{\text{P}_n}$ on top of a d.c.  drive $V_{\text{DC}}$, the probe excess Wigner distribution
is given by\cite{DegioWigner2013}: $\Delta W_{\text{P}_n}(t,\omega)=
-\frac{eV_{\text{P}_n}}{\hbar} \cos\left(2 \pi n f t + \phi\right) h_n\left(\omega
-\omega_{\text{DC}}\right)$, with $h_n(\omega)= \Big(f_{\text{eq}}(\omega-n\pi
f) - f_{\text{eq}}(\omega+n\pi f)\Big)/(2\pi nf)$ and $\omega_{\text{DC}}=-\text{e}
V_{\text{DC}}/\hbar$.  Inserting this expression for the probe Wigner
distribution in Eq.\eqref{SHOM} we can reconstruct the real and imaginary parts
of $\Delta W_{\text{S},n}$:
\begin{widetext}
\begin{subequations}
	\begin{align}
 \Re{(\widetilde{\Delta W}_{\text{S},n})} &=
 \frac{h}{8\text{e}^3V_{\text{P}_n}\mathcal{T}(1-\mathcal{T})} \left( \Delta
 S_{n,\phi=0}-\Delta S_{n,\phi=\pi} \right) = \int \md\omega \, \Re{\left(\Delta
 W_{\text{S},n}\left(\omega\right)\right)} \, h_n(\omega-\omega_{\text{DC}}) \label{n1r}  \\
 \Im{(\widetilde{\Delta W}_{\text{S},n})} &=
 \frac{h}{8\text{e}^3V_{\text{P}_n}\mathcal{T}(1-\mathcal{T})} \left(\Delta
 S_{n,\phi=\frac{\pi}{2}}-\Delta S_{n,\phi=\frac{3\pi}{2}}\right) = \int
 \md\omega \, \Im{\left(\Delta W_{\text{S},n}\left(\omega\right) \right) }\,
 h_n(\omega-\omega_{\text{DC}}) \label{n1i}
\end{align}
\end{subequations}
\end{widetext}
As in the $n=0$ case, the experimental signal is the convolution between $\Delta
W_{\text{S},n}$ and $h_n$.  The real and imaginary parts of $\Delta
W_{\text{S},n}$ are thus reconstructed using deconvolution techniques (see next
section).  A specific difficulty arises for the $n\neq 0$ terms, as their
reconstruction process requires the accurate knowledge of amplitude and phase of
the probe signals for various values of $n$.  The amplitude and phase
calibration of all the probe signals $V_{\text{P}_n}(t)$ is performed similarly to the
calibration of the amplitude and phase of the harmonics of the Lorentzian
voltage pulses.  As a result of the phase calibration, we find that, as
theoretically expected for the sine and Lorentzian drives,
$\widetilde{\Im{(\Delta W_{\text{S},n})}}=0$ for all $n$ and all $\omega$.
 Note that, as mentioned above, Coulomb interaction only modifies the phase and
 amplitude of sinusoidal drives.  As these are calibrated at the splitter
location, it means that we can simply ignore Coulomb interaction effects on the
probe signals.  Finally, when measuring the $n \neq 0$ harmonics, we also
systematically checked the linear dependence of the output noise with the probe
amplitude in order to check the validity of the linear approximation relating
$\Delta
W_{\text{P}_n}(t,\omega)$ to $V_{\text{P}_n}(t)$.

\subsection{Bayesian deconvolution method}

The relation between $\Delta W_{\text{S},n}(\omega)$ and $\widetilde{\Delta W}_{\text{S},n}(\omega)$ is
given by the convolution product, see Eqs.~\eqref{n0}, \eqref{n1r} and \eqref{n1i}:
\begin{subequations}
	\begin{align}
		\widetilde{\Delta W}_{\text{S},n}(\omega) &=
	\left(h_{n}\ast\Delta W_{\text{S},n}\right)(\omega)\\
		&=
	\int d\omega^{\prime}  h_{n}(\omega^{\prime}-\omega)\Delta W_{\text{S},n}(\omega^{\prime})
\label{eq:convolution}
\end{align}
\end{subequations}
In order to estimate $\Delta W_{\text{S},n}(\omega)$ based on
$\widetilde{\Delta W}_{\text{S},n}(\omega)$, we need to implement a
deconvolution algorithm. Since deconvolution is an ill-posed
problem, simply performing a division in Fourier space leads to an
estimation which is not robust to measurement errors. The
sensibility to errors, due to lost information, correspond to zero
or close to zero values of the Fourier transform of $h_{n}$. In
order to find a more robust estimation with correct physical
properties, we propose to add appropriate prior information on
$\Delta W_{\text{S},n}(\omega)$ thanks to a Bayesian framework
\cite{Ayasso2010,Djafari2015,Zhao2016}.
	
The discretized forward model for convolution \eqref{eq:convolution} can be expressed as
\begin{equation}
\widetilde{\mathbf{\Delta W}}_{\text{S},n} = \mathbf{H}_{n}.\mathbf{\Delta W}_{\text{S},n} + \mathbf{N}_{n},
\end{equation}
where bold characters stand for vectors and matrices resulting from
discretization. $\widetilde{\mathbf{\Delta W}}_{\text{S},n}$ is the
vector of data points, $\mathbf{H}_{n}$ is the convolution matrix,
and $\mathbf{\Delta W}_{\text{S},n}$ is the unknown quantity we are
looking for. The term $\mathbf{N}_{n}$ is added to take account for
all the errors (measurement and discretization). It is modeled as
Gaussian random vector, with known covariance matrix
$\mathbf{V}_{\text{e}}$ with diagonal
elements $V_{\text{e},i}$ estimated thanks to repeated experiments. This
gives the expression of the probability distribution of
$\widetilde{\mathbf{\Delta W}}_{\text{S},n}$ knowing $\mathbf{\Delta
W}_{\text{S},n}$ and $\mathbf{V}_{\text{e}}$, which is called the
likelihood:
\begin{widetext}
\begin{equation}
p\left(\widetilde{\mathbf{\Delta W}}_{\text{S},n}\left|\mathbf{\Delta W}_{\text{S},n},
	\mathbf{V}_{\text{e}} \right.\right)  \propto
	\exp\left(-\frac{1}{2} \left\| \widetilde{\mathbf{\Delta W}}_{\text{S},n}
	- \mathbf{H}_{n}.\mathbf{\Delta W}_{\text{S},n} \right\|^{2}_{\mathbf{V}_{\text{e}}}  \right)
\label{likelihood}
\end{equation}
\end{widetext}
where $\left\|\mathbf{x}\right\|_{\mathbf{V}_{\text{e}}}^{2}=\sum_i
\frac{x_{i}^2}{v_{\text{e}_{i}}}$. Finding the argument which maximizes the
likelihood, is equivalent to perform a division in Fourier space
since the convolution matrix $\mathbf{H}_{n}$ is diagonal in the
Fourier basis. This argument is dominated by
$\mathbf{H}^{-1}_{n}\mathbf{N}_{n}$ terms.
	
In the Bayesian framework, by adding a prior information, we want to
enforce physical properties such that $\Delta
W_{\text{S},n}\left(\omega\right) $ tends to zero when $\left|\omega
\right|$ increases. For this purpose, we assign a Gaussian prior
distribution on $\mathbf{\Delta W}_{\text{S},n}$:
\begin{equation}
p\left( \mathbf{\Delta W}_{\text{S},n}\Big| \mathbf{V}_{\text{f}} \right)  \propto
	\exp\left(
	-\frac{1}{2}\left\| \mathbf{\Delta W}_{\text{S},n} \right\|^{2}_{\mathbf{V}_{\text{f}}}
	\right).
\label{prior}
\end{equation}
For variances $\mathbf{V}_{\text{f}}$, we use the expression :
\begin{equation}
V_{\text{f}}(\omega) = v_{\text{f}}\exp \left( -\frac{\omega^{2}}{w^{2}} \right),
\label{eq:variances of delta w}
\end{equation}
where $v_\text{f}$ and $w$ are parameters tuned to enforce limit condition when
$\left|\omega \right|$ increases (the influence of the parameters $w$ and $v_{\text{f}}$
on the deconvoluted signal are presented in the Supplementary Note 4).
	
Applying Bayes'rule, the posterior probability distribution of $\mathbf{\Delta W}_{\text{S},n}$
combines likelihood \eqref{likelihood} and prior distribution \eqref{prior} :
\begin{widetext}
\begin{equation}
p\left(\mathbf{\Delta W}_{\text{S},n}\Big|\widetilde{\mathbf{\Delta
	W}}_{\text{S},n},\mathbf{V}_{\text{e}},\mathbf{V}_{\text{f}}\right) =
	\frac{p\left(\widetilde{\mathbf{\Delta
	W}}_{\text{S},n}\Big|\mathbf{\Delta
	W}_{\text{S},n},\mathbf{V}_{\text{e}}\right)p\left(\mathbf{\Delta
	W}_{\text{S},n}\Big|\mathbf{V}_{\text{f}}
	\right)}{p\left(\widetilde{\mathbf{\Delta
	W}}_{\text{S},n}\Big|\mathbf{V}_{\text{e}},\mathbf{V}_{\text{f}}\right)}\,.
\label{Bayes}
\end{equation}
\end{widetext}
The argument which maximizes this posterior distribution
\eqref{Bayes}, is the most likely estimate of $\mathbf{\Delta
W}_{\text{S},n}$ knowing both the measurement results
$\widetilde{\mathbf{\Delta W}}_{\text{S},n}$, $\mathbf{V}_{\text{e}}$, and
prior information encoded in $\mathbf{V}_{\text{f}}$. Indeed, this Maximum
A Posteriori (MAP), which also in this case is the Posterior Mean,
is robust to errors $\mathbf{N}_{n}$.
Because of model evidence, the term in the denominator of
\eqref{Bayes}, $p\left(\widetilde{\mathbf{\Delta
W}}_{\text{S},n}\Big|\mathbf{V}_{\text{e}},\mathbf{V}_{\text{f}}\right)$ does not
depend on $\mathbf{\Delta W}_{\text{S},n}$, MAP estimate becomes
equivalent to the minimization of the criterion:
\begin{widetext}
\begin{equation}
J\left(\mathbf{\Delta W}_{\text{S},n}\right) = \frac{1}{2}\left\|
	\widetilde{\mathbf{\Delta W}}_{\text{S},n}-\mathbf{H}_{n}.
	\mathbf{\Delta W}_{\text{S},n}  \right\|^{2}_{\mathbf{V}_{\text{e}}}+\frac{1}{2}
	\left\| \mathbf{\Delta W}_{\text{S},n}  \right\|^{2}_{\mathbf{V}_{\text{f}}}.
\label{eq:criterion}
\end{equation}
\end{widetext}
The estimated $\mathbf{\Delta W}_{\text{S},n}$ has to comply with a
box-constraint given by Pauli exclusion principle and
Cauchy-Schwartz inequality\cite{DegioWigner2013}. Consequently, the implemented
algorithm \ref{alg : Detailed algorithm} looks for a minimum of
criterion \eqref{eq:criterion} inside the box-constraint thanks to a
Projected Gradient Descent method \cite{Bertsekas1999}. In this
algorithm, $M_{n}(\omega)$ denotes Cauchy-Schwartz inequality
bounds $\forall n,\omega$ and is explicitly given in Ref.
\onlinecite{DegioWigner2013}.
	
\newpage
	
\begin{algorithm}[H]
	\caption{Detailed algorithm}
	\label{alg : Detailed algorithm}
	\begin{algorithmic}
			\STATE{Compute Cauchy-Schwartz bounds $M_{n}(\omega)$}
			\STATE{Choose amplitude $v_{\text{f}}$ and width $w$ of prior \eqref{eq:variances of delta w} for $V_{\text{f}}(\omega)$}		
			\STATE{Compute the minimum of criterion \eqref{eq:criterion}}
			\STATE{$\mathbf{\Delta W}_{\text{S},n} =
			\left(\mathbf{H}^{\top}\mathbf{V}^{-1}_{\text{e}}\mathbf{H}+
			 \mathbf{V}^{-1}_{\text{f}}\right)^{-1}\mathbf{H}^{\top}\mathbf{V}^{-1}_{\text{e}}\widetilde{\mathbf{\Delta
			W}}_{\text{S},n}$}
			\STATE{Project the solution inside the box given by Cauchy-Schwartz bounds:}
			\STATE{$\Delta W_{\text{S},n}(\omega) := \min(\Delta W_{\text{S},n}(\omega),M_{n}(\omega))$}
			\STATE{and $\Delta W_{\text{S},n}(\omega) := \max(\Delta W_{\text{S},n}(\omega),-M_{n}(\omega))$}
			\REPEAT
			\STATE{Compute the gradient of criterion \eqref{eq:criterion}}
			\STATE{
				$
				\nabla \left(\mathbf{ \Delta W}_{\text{S},n}\right)
					= -\mathbf{H}^{\top}\mathbf{V}^{-1}_{\text{e}}\left(\widetilde{\mathbf{\Delta W}}_{\text{S},n}
			-\mathbf{H}\mathbf{\Delta W}_{\text{S},n}\right)
					+\mathbf{V}^{-1}_{\text{f}}\mathbf{\Delta
					W}_{\text{S},n}
				$
				}
			\STATE{Project the gradient
			$\mathrm{\textbf{P}}\nabla\left(\mathbf{ \Delta W}_{\text{S},n}\right) $ to stay in the box-constraint}
			\IF{$|\Delta W_{\text{S},n}(\omega)| \geq M_{n}(\omega)$ and $\Delta W_{\text{S},n}(\omega)*\nabla\left( \Delta W_{\text{S},n}\right) (\omega) \leq 0$}
			\STATE{$\mathrm{P}\nabla\left( \Delta W_{\text{S},n}\right)(\omega) = 0$}
			\ELSE
			\STATE{$\mathrm{P}\nabla\left(\Delta W_{\text{S},n}\right)(\omega) = \nabla\left(\Delta W_{\text{S},n}\right)(\omega)$}
			\ENDIF
			\STATE{Compute the furthest displacement $d_{\infty}$
				in the box along $\mathrm{\textbf{P}}\nabla\left( \mathbf{\Delta W}_{\text{S},n} \right)$ direction}
			\FORALL{$\mathrm{P}\nabla\left(\Delta W_{\text{S},n}\right)(\omega) \neq 0$}
			\STATE{$d_{\infty} := \min\left(\frac{M_{n}(\omega)-\Delta W_{\text{S},n}(\omega)}{\mathrm{P}\nabla\left(\Delta W_{\text{S},n}\right)(\omega)},
			d_{\infty}\right)$
			}
			\ENDFOR
			\STATE{Compute the optimum displacement $d_{0}$ along
			$\mathrm{\textbf{P}}\nabla\left( \mathbf{\Delta W}_{\text{S},n}\right)$ direction}
			\STATE{$
				d_{0} = \left\|\mathrm{\textbf{P}}\nabla\left(
			\mathbf{\Delta W}_{\text{S},n}\right)\right\|^{-2}
				\left( \left\|\mathbf{H}\mathrm{\textbf{P}}\nabla\left(
				\mathbf{\Delta W}_{\text{S},n}\right)\right\|^{2}_{\mathbf{V}_{\text{e}}}
				\right.
				+\left.\left\|\mathrm{\textbf{P}}\nabla\left( \mathbf{\Delta
				W}_{\text{S},n}\right)\right\|^{2}_{\mathbf{V}_{\text{f}}}\right)
			$}
			\STATE{Compute one projected descent gradient step}
			\STATE{$\mathbf{\Delta W}_{\text{S},n} := \mathbf{\Delta W}_{\text{S},n}
			- \min(d_{0},d_{\infty})\mathrm{\textbf{P}}\nabla\left( \mathbf{\Delta W}_{\text{S},n}\right) $}			
			\UNTIL  $-\ln\left(p\left(\mathbf{\Delta W}_{\text{S},n}\left|
			\widetilde{\mathbf{\Delta W}}_{\text{S},n},\mathbf{V}_{\text{e}}, \mathbf{V}_{\text{f}}\right. \right) \right) $ is minimized \\
		\end{algorithmic}
\end{algorithm}

\pagebreak

\subsection{Electron and hole wavefunction extraction}

The extraction of electron and hole wavefunctions from the experimental
data for
$\Delta W_{\text{S}}(t,\omega)$ relies on an algorithm that recasts any excess
$T$-periodic single electron coherence under the
form given by Eq.~\eqref{eq:final-decomposition} of the article.  The algorithm
is a generalization of the Kahrunen-Lo\`eve analysis\cite{Book:Papoulis} to electron quantum
optics. It
is based on an exact diagonalization of the projections of
the single electron coherence (represented by $\Delta_0
W_{\text{S}}(t,\omega)$)
onto the electronic and hole quadrants with respect to the
reference chemical potential (here $\mu=0$). As explained in
Ref.~\onlinecite{DegioWigner2013}, these projections
are defined by decomposing the space of single particle state into
positive (for electrons) and negative energy (for holes) states.

In practice, they are obtained
through the following procedure:
after deconvolution, the experimental data come as a finite set of real values
$\Delta W_{\text{S},n}(\omega_k)$ where $\omega_k$ are the discretized values of
$\omega$ and $n=0,\pm 1, \pm 2 ...  \pm 5$.  First we add the thermal excess
$f_{eq}(\omega)-\Theta(-\omega)$ of the equilibrium Fermi Dirac distribution at
temperature $T_{\text{el}}$ to $\Delta W_{\text{S},n=0}(\omega)$ to obtain the
experimental dataset for $\Delta_0 W_{\text{S}}(\omega)$.  Then, in order to
extract a square matrix for the exact diagonalization, the next step is to
interpolate the data on a grid well suited to the electronic
and hole quadrants.  These two quadrants are defined in the frequency domain as
corresponding to the sectors where purely electronic (resp.  purely hole)
excitations contribute to $\Delta W_{\text{S},n}(\omega)$.  For a periodically
driven source, they correspond to $\omega \geq |n|\pi f$ for the
electron quadrant and to $\omega \leq -|n|\pi f$ for the hole quadrant
\cite{DegioWigner2013}.  For each $n$, the dataset
$\Delta_0 W_{\text{S},n}(\omega_k)$ is first interpolated using cubic splines to
infer a new dataset on a grid adapted to the electronic and hole quadrants (that
is such that this grid intersects the boundaries $\omega\pm n\pi f=0$ of the
electronic and hole quadrants). This new data grid has a discretization step
$\delta\omega$ such that $\hbar\,\delta\omega\simeq \SI{0.19}{\micro\eV}$.
This
dataset is then used to build the matrices corresponding to the
projections on the electron and hole quadrants of
this interpolated data for $\Delta_0 W_{\text{S},n}(\omega)$.

Due to time periodicity of the single electron Wigner function
$\Delta_0 W_{\text{S}}(t,\omega)=\Delta_0 W_{\text{S}}(t+T,\omega)$,
diagonalizing these two projections
onto the electron and hole quadrants leads to
electronic ($\alpha=\text{e}$) and hole ($\alpha=\text{h}$) probability spectral bands $g_i^{(\alpha \alpha)}(\nu)$
($i$ being a band index) depending on
quasi-pulsation interval $0\leq \nu\leq 2\pi f$ associated with the time period
$T$. The corresponding eigenvectors are the Floquet version of Bloch waves of
solid state physics.

Then, the electron and hole wavefunctions
$\varphi^{(\alpha)}_{l,i}$ generated at each time period\cite{These:Roussel,Roussel:2016-2}
are the analogous of the Wannier functions\cite{Wannier:1937-1}.
They consist of normalized single-particle wavepackets such that
\begin{subequations}
	\begin{align}
		\varphi^{(\alpha)}_{l,i}(t)&=\varphi^{(\alpha)}_i(t-lT) \\
		\langle \varphi^{(\alpha')}_{l',i'}|\varphi^{(\alpha)}_{l,i}\rangle &=
		\delta_{i,i'}\,\delta_{\alpha,\alpha'}\,\delta_{l,l'}\,.
	\end{align}
\end{subequations}
where $\alpha$ denotes the electron or hole label and $i$
the band index. These electron and hole wavefunctions
are therefore very well suited to describe the excitations generated by time periodic
electron beams.
In solid state physics\cite{Marzari:2012-1}, the Wannier wavefunctions are not
uniquely defined since one can impose an arbitrary quasi-momentum
dependent
phase in front of each Bloch wave. Here the same problem is present and,
exactly as in solid state physics, this ambiguity is
lifted by minimizing their time spreading.
This provides us electronic and hole atoms of signals that are maximally localized
 in the time domain.
Finally, the electron
coherence $g^{(\text{ee})}_i(l)$
between electronic atoms of signals translated by $l\in\mathbb{Z}^*$
time periods as well as the hole coherence $g^{(\text{hh})}_i(l)$
can be obtained from the probability
spectra through
Fourier transform. For $l\in \mathbb{Z}$:
\begin{equation}
g^{(\alpha \alpha)}_i(l)=\int_0^{2\pi f} e^{i \nu
l/f}g^{(\alpha)}_i(\nu)\frac{d\nu}{2\pi f}
\end{equation} .
Note that there is no electronic coherence between atoms of signal
of the same type of excitation but with different band index.
The electron/hole
coherences $g^{(\text{eh})}_{i,i'}(l)$ are defined as the single electron
coherence between the electronic atom of signal $\varphi^{(\text{e})}_{i,l}$
and the hole atom of signal $\varphi^{(\text{h})}_{i',0}$~:
\begin{eqnarray}
	g^{(\text{eh})}_{i,i'}(l)& = & \mathrm{Tr}\left(\psi[\varphi^{(\text{e})}_{l,i}]\,\rho\,
	\psi^\dagger[\varphi^{(\text{h})}_{0,i'}]\right) \\
	& = & \int \md t\, \md t' \varphi^{(\text{e})}_{l,i}(t)^* \; \Delta_0
	\mathcal{G}(t,t') \; \varphi^{(\text{h})}_{0,i'}(t')\,.
\end{eqnarray}
where $l\in\mathbb{Z}$, $i$ and $i'$ are possibly
different. It is obtained from the electronic Wigner function
using the explicit numerical data for the electronic
and hole wavefunctions.

\subsection{Purity indicator}

The general expression of the state purity is beyond the scope of this paper.
We focus here on the two limiting cases considered in this paper: the sinusoidal
drives where one electron and one hole wavefunction are generated and the
periodic train of single electron Lorentzian pulses where two electronic
wavefunctions need to be considered (the probability for hole emission can be
neglected).

Let us consider first the sinusoidal drive case with only one electron
and one hole branch so that the branch index $i$ can be dropped out.
Time periodicity implies that
only coherences between Floquet-Bloch eigenvectors with the same quasi-pulsation
do not vanish.  As only one electron and one hole wavepackets are emitted, we
can therefore consider, at each given quasi-pulsation $0\leq \nu < 2\pi f$, the
reduced density matrix $\rho_{\text{eh}}$ in the occupation number basis of the
electron and hole states: $|n_\text{e} n_\text{h}\rangle$.  $n_\text{e}=0$ or $1$ and $n_\text{h}=0$ or $1$
are the occupation numbers of the corresponding single particle states.  As a
result of the superselection rule that forbids quantum superposition between
states with fermion numbers of different parity \cite{Wick1952,Johansson2016},
$\rho_{\text{eh}}$ corresponds to a pure state
in three situations: either the electronic and hole levels are both filled
(state $|11\rangle$), or both empty (state $|00\rangle$), or populated in a
coherent way (state $u|01\rangle + v|10\rangle$ with $|u|^2+|v|^2=1$).  Any
deviation from purity thus reflects incoherent electron/hole processes.  The
purity indicator which is defined as $\mathrm{Tr}(\rho_{\text{eh}}^2)$ is a good
quantity for measuring the weight of coherent processes: \begin{equation}
\mathrm{Tr}(\rho_{\text{eh}}^2) = 1-2A(\nu)(1-A(\nu))-2B(\nu)(1-B(\nu)) \end{equation}
where \begin{align}
	A(\nu) &= g^{(\text{ee})}(\nu)(1-g^{(\text{hh})}(\nu))-|g^{(\text{eh})}(\nu)|^2 \\ B(\nu) &=
	g^{(\text{hh})}(\nu)(1-g^{(\text{ee})}(\nu))-|g^{(\text{eh})}(\nu)|^2
\end{align} are computed in terms of the eigenvalues $g^{(\text{ee})}(\nu)$ and
$g^{(\text{hh})}(\nu)$ obtained from our diagonalization algorithm and of the
corresponding electron/hole coherences $g^{(\text{eh})}(\nu)$.

Let us now discuss all the reduced density matrices $\rho_{\text{eh}}$ for all $0\leq
\nu<2\pi f$.  When Wick's theorem is valid, there are no correlations between
different quasi-pulsations $\nu_1\neq \nu_2$.  Consequently, we can take the
infinite-dimensional product over all the Floquet-Bloch pairs of electron and
hole modes for $0\leq\nu< 2\pi f$ and take the trace of its square which is a
formal infinite product over all $0\leq \nu <2\pi f$.  This quantity has to be
regularized in the infrared by discretizing the quasi energies $\nu_n:2\pi f
n/N$ for $n=0\ldots N-1$ and taking the $1/N$-th power of the result.  This
procedure leads to the resulting quantity
\begin{widetext}
\begin{equation}
	\label{purity}
	\mathbb{P} = \exp\left[
		\int_0^{2\pi f} \ln\left( 1-2A(\nu)(1-A(\nu))-2B(\nu)(1-B(\nu))
		\right) \frac{d\nu}{2\pi f} \right]
\end{equation}
\end{widetext}
which is equal to unity if and only if the many body state
is pure and obtained from a Fermi sea vacuum by adding on top of it coherent
superposition of electron/hole pairs. When there are incoherent processes such
as in the case of non-zero temperature, $\mathbb{P}<1$.  In the present
situation, the condition $A(\nu)=B(\nu)=0$ which
ensures unit purity corresponds to
$|g^{(\text{eh})}(\nu)|^2=g^{(\text{ee})}(\nu)(1-g^{(\text{hh})}(\nu))=g^{(\text{hh})}(\nu)(1-g^{(\text{ee})}(\nu))$,
which implies that $g^{(\text{ee})}(\nu)=g^{(\text{hh})}(\nu)$.  Then, the condition on
$g^{(\text{eh})}(\nu)$ ensures that, for each quasi-pulsation $\nu$, we have acted on
the Fermi sea $|F\rangle$ through the coherent sum of the identity operator, and
of the elementary electron/hole pair creation operator that is the product of a
creation operator for the electron single particle state and of a destruction
operator for the hole single particle state.  This is equivalent to putting each
quasi-particle which, in $|F\rangle$, is in the hole state $\varphi^{(\text{h})}_\nu$,
into the linear combination $u(\nu)\varphi_\nu^{(\text{h})}+ v(\nu)\varphi_\nu^{(\text{e})}$.
The resulting many-body state is then pure and of the form:
\begin{equation}
|\Psi\rangle= \prod_{0\leq \nu<2\pi f} \left(u(\nu)+
	 v(\nu)\psi^\dagger[\varphi_\nu^{(\text{e})}]\psi[\varphi_\nu^{(\text{h})}]\right)|F\rangle \,.
\end{equation}
This specific form was also obtained in Ref. \onlinecite{Yin2019} which considered a conductor at zero temperature described by a single particle time-dependent scattering matrix.  It reduces to the form given by Vanevic \textit{et
al.}\cite{Vanevic2016} in the case where the Floquet-Bloch spectrum is flat as a
function of $\nu$ (in which case there are no interperiod electronic and hole
coherences).

Eq.~\eqref{purity} for the purity can be adapted to the Lorentzian case we study
in the paper. Three electronic bands need to be considered at most for the
Lorentzian pulse carrying two electrons.  These bands are not coupled in our
specific experimental situation (the term coupling the two bands are the
electron-hole coherences $g^{(\text{eh})}_{ij}$ which we measure to be negligible).  In
this case, the extension of Eq.~\eqref{purity} is straightforward:
\begin{subequations}
\begin{align}
	\mathbb{P} &= \mathbb{P}_1 \times \mathbb{P}_2 \times \mathbb{P}_3 \\
\mathbb{P}_i &= \exp\left[
		\int_0^{2\pi f} \ln\left( 1-2g^{(\text{ee})}_i(\nu)(1-g^{(\text{ee})}_i(\nu))
		\right) \frac{d\nu}{2\pi f} \right]
\end{align}
\end{subequations}
where we have used the simplified expressions of $A_i(\nu)=
g^{(\text{ee})}_i(\nu)$ and $B_i(\nu)=0$ in the case where $g^{(\text{hh})}_i(\nu) \approx
g^{(\text{eh})}_{ij}(\nu) \approx 0$.  The pure state $\mathbb{P}=\mathbb{P}_i=1$ is
only recovered for $g^{(\text{ee})}_i(\nu)=0$ or $1$.  In our experimental situation for
the single electron Lorentzian pulse, two wavefunctions are emitted with
probabilities smaller than 1 and we find $\mathbb{P}=0.75$, showing that the
generated state is a mixture of two single single electron wavefunctions.  For
the two-electrons Lorentzian pulse, we find a slightly smaller purity
$\mathbb{P}=0.68$.

\section*{Data Availability}
The data that support the findings of this study are available from the corresponding author (G.F.) upon reasonable request.

\section*{Acknowledgements} This work has been supported by ANR grants '1shot
reloaded' (ANR-14-CE32-0017), the project EMPIR 17FUN04 SEQUOIA, and the ERC
consolidator grant 'EQuO' (No.  648236).

\section*{Author Contributions}

YJ fabricated the sample on GaAs/AlGaAs heterostructures grown by AC and UG. RB, AM, and MK conducted the measurements. Deconvolution algorithms were developed by CCh, RB and AMD. The algorithm for the extraction of electron and hole wavefunctions was developed by BR, CCa and PD. RB, AM, MK, JMB, EB, BP and GF participated to the data analysis. All authors participated to the writing of the manuscript. GF supervised the project.

\section*{Competing Interests}
The authors declare no competing interests.

\newpage

\newpage

\newpage

\newpage

\newpage

\newpage

\newpage

\end{document}